\documentclass[a4paper,11pt]{article}
\pdfoutput=1 % if your are submitting a pdflatex (i.e. if you have
\usepackage{jheppub} % for details on the use of the package, please
\usepackage{bigints}
\usepackage{graphicx}
\usepackage{subfigure}
\usepackage{placeins}
\usepackage{stackengine}
\usepackage{array}
\usepackage{enumerate}
\allowdisplaybreaks

\newcolumntype{M}[1]{>{\centering\arraybackslash}m{#1}}
\newcolumntype{N}{@{}m{0pt}@{}}
\usepackage[T1]{fontenc} % if needed

\title{\boldmath Massive vector particle tunneling from Kerr-Newman-de Sitter black hole under generalized uncertainty principle}

\author[a]{Yenshembam Priyobarta Singh,\note{Corresponding author.}}
\author[a,1]{Telem Ibungochouba Singh}
\affiliation[a]{Department of Mathematics,\\Manipur University, Canchipur, Imphal, Manipur, India }
\emailAdd{priyoyensh@gmail.com}
\emailAdd{ibungochouba@rediffmail.com}
\abstract{The quantum tunneling of charged massive vector boson particles across the event horizon of Kerr-Newman-de Sitter black hole is investigated under the influence of quantum gravity effects. \textbf{ The modified Hawking temperatures and heat capacities  across the event horizon of KNdS black hole are derived in 3-dimensional and 4-dimensional frame dragging coordinates. It is found that due to quantum gravity effects the modified Hawking temperatures and heat capacities depend on the mass and angular momentum of the emitted vector boson particles. For 3-dimensional KNdS black hole, the modified Hawking temperature is lower than the original Hawking temperature but the modified heat capacity is higher than the original heat capacity due to quantum gravity effects. In the case of 4-dimensional KNdS black hole, the modified Hawking temperature and heat capacity are lower or greater than the original Hawking temperature and heat capacity depending upon the choices of black hole parameters due to quantum gravity effects.} We also discuss the remnant and graphical analysis of the modified Hawking temperatures and heat capacities.}

\begin{document} 
\maketitle
\flushbottom

\section{Introduction}
\label{sec:intro}
\par In the early 1970s, Hawking proposed a black hole radiation called Hawking radiation using quantum field theory techniques on a curve space-time background \cite{hawking1,hawking2}. The discovery of Hawking radiation is one of the significant developments in gravitational physics. It shows that black holes have a thermodynamical features. Refs. \cite{bekenstein1,bekenstein2} determined that the horizon area and the black hole's entropy are proportional. More precisely, the black hole's entropy is equal to one-fourth of its horizon area. Since then, theoretical physicists have paid a lot of attention to Hawking radiation and a number of approaches have been used to determine Hawking radiation.

\par Generally, there are two approaches to investigate the imaginary part of the action. They are the radial null geodesic method and the Hamilton-Jacobi method. The radial null geodesic method was put forth by Parikh and Wilczek \cite{kraus1, kraus2, kraus3,parikh1}. In this method, the emitted particles act as the potential barrier and the semiclassical WKB approximation is used to determine the imaginary part of the radial action. Later, Zhang and Zhao \cite{zhang1, zhang2, zhang3} have made significant progress by extending Parikh-Wilczek method to non-spherical symmetric stationary black holes and the radiation of charged massive particle. \textbf{Moreover, using the Parikh-Wilczek method, extensive investigations of many types of black holes have been conducted  \cite{vagenas2, vagenas3,v4,vagenas4,a3}}.  The second is the Hamilton-Jacobi method, used by Angheben et al. \cite{angheben1} and Vagenas \cite{vagenas1} as an extension of the complex path integral method introduced by Padmanabhan et al. \cite{padmanabhan1, padmanabhan2, padmanabhan3}. In both methods, the WKB approximation is used to investigate the tunneling probability for classically forbidden trajectory from inside to outside the horizon and it is given by $\Gamma=\exp \left[	-\frac{2}{\hbar} \text{Im}~ I \right]$, where $I$ and $\hbar$ denote the semiclassical action of the outgoing particle and Planck's constant. Kerner and Mann \cite{kerner1, kerner2} investigated the tunneling of spin-1/2 fermions particle from rotating and non-rotating black holes using Dirac equation, Pauli sigma matrix and Fenyman prescription. By  the fermions tunneling method the Hawking radiation from the black ring is also derived \cite{jiang1}. Using their method, many interesting results have been obtained in \cite{ ren1,wang1,rahaman1,ibungochouba1}. Damour and Ruffini \cite{damour1} used a new approach called tortoise coordinate transformation  to study Hawking radiation.  Later, Sannan \cite{sannan1} extended the work of Damour and Ruffini by deriving the probability distributions for boson and fermions emission from a black hole.\textbf{ Moreover, topological approach has been used to derive Hawking temperature of black holes \cite{robson1, robson2, ali11, ali12, ali13}. }

The existence of a minimum measurable length that can be identified with the order of the Planck scale is predicted by numerous quantum gravity theories, such as string theory, loop quantum gravity, non-commutative geometry and  Gedanken experiments \cite{townsend1, yoneya1, konishi1, maggiore1, garay1, scardigli1, amelino1}. Constructing new theoretical models is one of the fields of study for quantum gravity. The Generalized Uncertainty Principle (GUP) is a modified theory that realizes this minimum length. It is the generalization of the Heisenberg uncertainty principle (HUP). The modified fundamental commutation relation proposed by Kempf et al. \cite{kempf1} is of the form
\begin{align}
\bigl[x_i,p_j \bigr]=i \hbar \delta_{i,j} \bigl[1+\beta p^2 \bigr],
\end{align} 
where the positions $x_i $ and momentum operators $p_j$ are defined by
\begin{align}
x_i=x_{0i}, ~~~~~ p_j=p_{0j} \bigl( 1+\beta p_{0}^2 \bigr).
\end{align}
Then $x_{0i}$ and $p_{0j}$ satisfy the canonical commutation relations  as $\bigl[x_{0i},p_{0j}\bigr]=i \hbar ~\delta_{ij}$. The corresponding GUP takes the form
\begin{align}
\Delta x \Delta p \geq \frac{\hbar}{2} \bigl[ 1+  (\Delta p)^2 \beta \bigr],
\end{align}
where $\beta=\frac{\beta_0 l^{2}_p}{\hbar^2}=\frac{\beta_0}{M_{p}^2 c^2}$. Here $\beta_0 (\leq 10^{34})$, $l_p$ and $M_p$ represent the dimensionless parameter of order unity, Planck length and Planck mass respectively.

The implications of the aspects of GUP have been investigated in many contexts such as modifications of the quantum Hall effect, unruh effect, neutrino oscillations, Landau levels, Newton's law, cosmology, and the weak equivalence principle (WEP) \cite{das1, ali1, zhu1, sprenger1, tawfik1, ali2, vagnozzi,s2,s3}. Adler et al. \cite{adler1} studied the influence of GUP on the thermodynamics of the Schwarzschild black hole. \textbf{Later, Ref. \cite{ovgunR} investigated the tunneling of scalar and fermion particles from a Schwarzschild black hole immersed in an electromagnetic universe under the effect of GUP} . Further, using the Parikh-Wilczek tunnelling approach, Nozari and Saghaf \cite{nozari1} studied the Hawking radiation for massless scalar particles in the background geometry of Schwarzschild black hole and retrieved the tunnelling rate as well as the corrected Hawking temperature by taking GUP into account. It should be noted that  the GUP has also influenced the thermodynamics of black holes.  As a result, the GUP concept has been applied to various black holes in order to study their thermodynamics properties  \cite{a2}. Applying the WKB approximation to the Dirac equation, the tunnelling of fermions from the Kerr and Kerr-Newman black holes are studied in \cite{li1,jian1}. Yale \cite{yale1} investigated the tunneling of arbitrary scalars, fermions and spin-1 bosons by ignoring back-reaction effects. Banerjee and Majhi \cite{banerjee1,banerjee2} discussed the tunneling of fermions and bosons at the event horizon of black holes beyond the semiclassical approximation. They derived the modified Hawking temperature and change in black hole entropy. Ibungochouba et al. \cite{ibungochouba2} studied fermions quantum tunneling from the BTZ black hole. Li \cite{li2} studied the tunneling of massive spin-1 particle from Reissner-Nordstrom and Kerr black hole under the effect of quantum gravity. \"Ovg\"un et al. \cite{ovgun1} investigated the charged massive bosons tunneling from noncommutative charged black holes.

\textbf{GUP prevents a black hole from complete evaporation as the black hole's mass approaches the Planck scale. This late stage of a black hole under Hawking evaporation is referred as black hole remnant. A black hole remnant consists of a black hole phase that evaporates under the Hawking radiation, which is either stable or long-lived/meta-stable. Many researchers have investigated the black hole remnant in \cite{adler,adler2,chen1, chen2, myung1, gangopadhyay1, feng1,eslamR}. In the context of the information loss paradox, the study of black hole remnant plays a crucial role \cite{unruh, chenR}.}

%Many studies have shown that, GUP avoid the black hole to evaporate completely. The black hole has a phase transition close to the Planck scale and reaches thermal equilibrium with its surroundings. As a result, the black hole's evaporation stops, leaving behind a stable remnant \cite{chen1, chen2, myung1, gangopadhyay1, feng1}. Li \cite{li2} studied the tunneling of massive spin-1 particle from Reissner-Nordstrom and Kerr black hole under the effect of quantum gravity. Ovgun et al. \cite{ovgun1} investigated the charged massive bosons tunneling from noncommutative charged black holes.

%In this paper, the tunneling of massive vector bosons particles across the event horizon of the KNdS black hole is investigated. 

This article's layout is constructed in the following manner: In section 2, we revisit the generalized field equation for massive vector boson particles. The quantum tunneling of massive vector boson from KNdS black hole under quantum gravity effects is presented in section 3. Further, section 4 provides the analysis of the remnant of 3-dimensional KNdS black hole under quantum gravity effects. The graphical analysis of quantum corrected Hawking temperatures and heat capacities are presented in section 5. Lastly, some concluding remarks are presented in section 6.
\section{Generalized field equations for massive vector bosons}

The GUP-corrected Lagrangian of massive vector field $\Psi_\mu$ is given by \cite{li2}
\begin{align} \label{eqn 1}
\mathfrak{L}_{GUP}=& -\frac{1}{2} \left(\mathfrak{D}^{+}_\mu \Psi^{+}_\nu-\mathfrak{D}^{+}_\nu \Psi^{+}_\mu	\right) \left(\mathfrak{D}^{-\mu} \Psi^{-\nu}-\mathfrak{D}^{-\nu} \Psi^{-\mu} 	\right)-\dfrac{m^2}{\hbar^2} W^{+}_{\mu} W^{-\mu} 	\nonumber \\
& -\dfrac{i}{\hbar} e F^{\mu \nu} W^{+}_{\mu} W^{-}_{\nu},
\end{align}
%Here $\mathfrak{D}_{0}^{\pm}=\left(1+\beta \hbar^2 g^{00} \mathcal{D}_{0}^{\pm 2}	\right)\mathcal{D}_{0}^{\pm}$, $\mathfrak{D}_{i}^{\pm}=\left(1-\beta \hbar^2 g^{ii} \mathcal{D}_{i}^{\pm 2}	\right)\mathcal{D}_{i}^{\pm}$ with $D_{\mu}^{\pm}=\nabla_{\mu} \pm \frac{i}{\hbar} e A_{\mu}$, $\mathfrak{D}$
where $\mathcal{D}_{0}^{\pm}=\left(1+\beta \hbar^2 g^{00} D_{0}^{\pm 2}	\right)D_{0}^{\pm}$, $\mathcal{D}_{i}^{\pm}=\left(1-\beta \hbar^2 g^{ii} D_{i}^{\pm 2}	\right)D_{i}^{\pm}$,\\ $F_{\mu \nu}=\widehat{\nabla}_{\mu}A_{\nu}-\widehat{\nabla}_{\nu}A_{\mu}$ with $D_{\mu}^{\pm}=\nabla_{\mu} \pm \frac{i}{\hbar} e A_{\mu}$, $\widehat{\nabla}_{0}=\left(1+\beta \hbar^2 g^{00} \nabla_{0}^2 \right) \nabla_{0}$ and  $\widehat{\nabla}_{i}=\left(1-\beta \hbar^2 g^{ii} \nabla_{i}^2 \right) \nabla_{i}$. Here $e$, $m$ and $A_\mu$ denote the charge of $W^+$ boson, mass of the vector particle and electromagnetic potential of the black hole respectively.

The generalized action of the massive vector bosons particles takes the form
\begin{align} \label{eqn 2}
\mathcal{S}_{GUP}=  \displaystyle\int dx^4 \sqrt{-g}~ \mathfrak{L}_{GUP} \left( \Psi_{\mu}^{\pm}, \partial_{\mu}	\Psi_{\nu}^{\pm}, \partial_{\mu} \partial_{\rho} \Psi_{\nu}^{\pm}, \partial_{\mu} \partial_{\rho} \partial_{\lambda} \Psi_{\nu}^{\pm}	\right).
\end{align}

Following from eq. \eqref{eqn 2}, the modified wave equation for massive vector bosons is obtained as 
\begin{align} \label{eqn 3}
\partial_{\mu} & (\sqrt{-g} \Psi^{\mu \nu})-3 \beta \partial_{0} \left[ \sqrt{-g} {}{}g^{00} \left( e^2 A_{0}^2+i \hbar e \nabla_{0} A_0	\right) \Psi^{0v}	\right] \nonumber \\
& +3 \beta ~\partial_{i} \left[ \sqrt{-g} {}{}g^{ii} \left( e^2 A_{i}^2+i \hbar e \nabla_{i} A_i	\right) \Psi^{iv}	\right]+ 3 \beta \partial_0 \partial_0 \left[ \sqrt{-g} {}{}g^{00} i \hbar e A_0 \Psi^{0v} \right] \nonumber \\
& -3 \beta \partial_i \partial_i \left[ \sqrt{-g} {}{}g^{ii} i \hbar e A_i \Psi^{iv} \right]+ \beta \hbar^2 \partial_0 \partial_0 \partial_0 \left[ \sqrt{-g} {}{}g^{00} \Psi^{0v} \right]  \nonumber \\
&- \beta \hbar^2 \partial_i \partial_i \partial_i \left[ \sqrt{-g} {}{}g^{ii} \Psi^{iv} \right] + \sqrt{-g} \frac{i}{\hbar} e A_\mu \Psi^{\mu \nu}- \sqrt{-g} \frac{m^2}{\hbar^2}  \Psi^{\nu} \nonumber \\
& -\sqrt{-g} \frac{i}{\hbar} e F^{\mu \nu}  \Psi_{\mu} + \beta \sqrt{-g} g^{00} \left[ i \hbar e \nabla_{0} \nabla_{0} A_0+3 e^2 A_0 \nabla_{0} A_0-\frac{i}{\hbar} e^3 A_{0}^{3} \right] \Psi^{0 \nu}	\nonumber \\
& - \beta \sqrt{-g} g^{ii} \left[ i \hbar e \nabla_{i} \nabla_{i} A_i+3 e^2 A_i \nabla_{i} A_i-\frac{i}{\hbar} e^3 A_{i}^{3} \right] \Psi^{i \nu} ~=~ 0,
\end{align}
where GUP modified anti-symmetric tensor $\Psi_{\mu \nu}$ is given by  $\Psi_{\mu \nu}=\mathcal{D}_\mu \Psi_\nu- \mathcal{D}_\nu \Psi_\mu$.
\section{Quantum tunneling from KNdS black hole}
The line element of KNdS black hole in well known Boyer-Lindquist coordinates $(t,r,\theta,\varphi)$ \cite{carter1} is expressed as

\begin{eqnarray} \label{eqn 4}
ds^2&=&-\left(\dfrac{\Delta- \Delta_{\theta} ~ a^2\sin^2\theta}{ \Sigma^2 \rho^2}\right)dt^2-\dfrac{2a~  \sin^2\theta [(r^2+a^2) \Delta_{\theta}-\Delta]}{ \Sigma^2 \rho^2}dt d\varphi +\dfrac{\rho^2}{\Delta}dr^2 \cr &&+\dfrac{\rho^2}{\Delta_{\theta}} d\theta^2 +\dfrac{\sin^2\theta \left[ (r^2+a^2)^2~ \Delta_{\theta}-\Delta a^2\sin^2\theta \right]}{\Sigma^2 \rho^2}  d\varphi^2,
\end{eqnarray}

where 
\begin{align*}
&\Sigma=1+\dfrac{\Lambda a^2}{3},~ \rho^2=r^2+a^2\cos^2\theta,~ \Delta_{\theta}=1+\dfrac{\Lambda a^2 \cos^2\theta}{3},\\
&\Delta=\left(1-\dfrac{\Lambda a^2}{3}	\right) \left( r^2+a^2\right)-2Mr+Q^2.
\end{align*}
%The electromagnetic potential is given by $(A_0,A_1,A_2,A_3)$ where
%$A_0=$

The electromagnetic potential $A_\mu$ of KNdS black hole is given by 
\begin{align} \label{eqn 5}
A_{\mu}=\dfrac{Q r \left( \delta_{\mu}^{t}-a\sin^2 \theta \delta_{\mu}^{\varphi}	\right)}{\rho^2 \Sigma}.
\end{align}
Eq. \eqref{eqn 4} describes a charged rotating black hole with mass $m$, spin parameter $a$ and charge $Q$ with the cosmological constant $\Lambda$. The case $\Lambda=0$ gives the solution of the Kerr-Newman black hole. For $\Lambda>0$ or $\Lambda<0$, eq. \eqref{eqn 4} represents KNdS black hole or anti KNdS black hole respectively.  
%taken as  $(A_0,0,0,A_3)$ where\\ $A_{0}=\dfrac{-(r^2+a^2) Q r}{(r^2+a^2)^2-\Delta a^2\sin^2\theta}$.
%\\
%The electromagnetic potential is given as $(A_{t},0,0,A_{\varphi})$ where
%$A_{t}=\dfrac{-Q r}{\Sigma}$ and $A_{\varphi}=\dfrac{Q r a \sin^2 \theta}{\Sigma}$.
%\\

The horizons of the KNdS can be obtained from the event horizon equation as
\begin{equation}\label{eqn 6}
\Delta=\left(1-\dfrac{\Lambda r^2}{3}	\right) \left( r^2+a^2\right)-2Mr+Q^2=0.
\end{equation}
%\frac{1}{3} \Lambda \left(r-r_{C}	\right)	\left(r-r_{H}	\right)	\left(r-r_{+}	\right)	\left(r-r_{-}	\right)=0
%\begin{equation} \label{eqn 303}
%\Delta=\left(1-\dfrac{\Lambda a^2}{3}	\right) \left( r^2+a^2\right)-2Mr+Q^2=0.
%\end{equation}

If $\Lambda>0$, Eq. \eqref{eqn 6} gives four real roots whenever $\frac{1}{\Lambda}\gg M^2 > Q^2+a^2$. The four roots are $r_C$, $r_H$, $r_+$ and $r_-$ $\left(r_C> r_H>r_{+}>r_{-}\right)$. Here $r_C$ denotes the cosmological horizon, $r_H$ represents the event horizon and $r_{-}$ corresponds to the Cauchy horizon. One reaches singularity $r=0$, $\theta=\frac{\pi}{2}$ and on other side of $r=0$, $r=r_-$ is considered as another cosmological horizon \cite{gibbons1}.

In this coordinate system, the event horizon and the infinite red-shift surface do not coincide due to rotation. Because of this, it is inconvenient to study the characteristic of tunneling radiation.
So, we perform the dragging coordinate transformation \cite{zhangC}
\begin{equation}\label{eqn 7}
\dfrac{d \varphi}{dt}=\Omega=\dfrac{-g_{t \varphi}}{g_{\varphi \varphi}}.
\end{equation}

The line element \eqref{eqn 4} is reduced to
\begin{eqnarray} \label{eqn 8}
ds^2=-\dfrac{ \Delta \Delta_{\theta} \rho^2}{ \Sigma^2 \left[ \Delta_\theta(r^2+a^2)^2-\Delta a^2\sin^2\theta \right]} dt^2 +\dfrac{\rho^2}{\Delta}dr^2+ \dfrac{\rho^2}{\Delta_\theta} d\theta^2.
\end{eqnarray}

The non-vanishing component of electromagnetic potential $A_0$ is given by
\begin{align}
A_0=\dfrac{(r^2+a^2)\Delta_\theta Q r}{\Sigma \left[\Delta_\theta (r^2+a^2)^2-\Delta a^2 \sin^2\theta \right]}.
\end{align}

The dragging coordinate transformation makes the geometrical optical limit a reliable approximation; hence, the WKB approximation can be applied.

Near the event horizon $r=r_H$, we use the approximation 
\begin{align} \label{eqn 10}
\Delta(r)=& \Delta(r_H)+(r-r_H) \Delta,_r(r_H)+O\left((r-r_H)^2\right) \nonumber \\
& \approx (r-r_H) \Delta,_r(r_H),
\end{align}
where $r_H$ is defined as
\begin{align} \label{eq r}
r_{H}= \dfrac{1}{\alpha_1} \left( 1+\dfrac{4 \Lambda M^2}{3 \beta_{1}^2 \alpha_1} + \dots	\right) \left( M+ \sqrt{M^2-(a^2+Q^2)\alpha_1}	\right)	,
\end{align}
where $\beta=\sqrt{1-\frac{\Lambda a^2}{3}}$ and $\alpha_{1}=\sqrt{\left(1+ \frac{\Lambda a^2}{3} \right)^2+\frac{4 \Lambda Q^2}{3}}$.
Then eq. \eqref{eqn 8} takes the form
\begin{align}\label{eqn 11}
ds^2=-\dfrac{(r-r_H) \Delta,_r(r_H)  ~\rho^2(r_H)}{\Sigma^2  \left(r^{2}_{H}+a^2\right)^2} dt^2+\dfrac{\rho^2(r_H)}{(r-r_H) \Delta,_r(r_H)} dr^2+\dfrac{\rho^2(r_H)}{\Delta_\theta}d\theta^2.
\end{align}

%At the event horizon, the non-zero component of the electromagnetic potential is
%\begin{align} \label{eqn 12}
%A_0=\dfrac{Q r_H \Delta_\theta}{\Sigma \Delta_\theta \left(r^{2}_{H}+a^2\right)}.
%\end{align}

The metric determinant and contravariant components of the line element \eqref{eqn 11}  are as follows
\begin{align}
& g= -\dfrac{\rho^6 (r_H)}{\Sigma^2 \left(r_{H}^2+a^2\right)^2 \Delta_{\theta}}, \nonumber \\
& g^{00}=-\dfrac{\Sigma^2  \left(r^{2}_{H}+a^2\right)^2}{(r-r_H) \Delta,_r(r_H)  ~\rho^2(r_H)}, \nonumber\\
& g^{11}=\dfrac{(r-r_H) \Delta,_r(r_H)}{\rho^2(r_H)},~~ g^{22}=\dfrac{\Delta_\theta}{\rho^2(r_H)}, \nonumber\\
& g^{01}=g^{02}=g^{10}=g^{12}=g^{20}=g^{21}=0.
\end{align}

 The angular velocity at the event horizon is given as
 \begin{align}
 \Omega=\dfrac{a}{r^{2}_h+a^2}.
 \end{align}
The surface gravity near the event horizon of the KNdS black hole is obtained as
\begin{align}
\kappa=& \lim_{g_{00}\rightarrow 0} \Biggl(-\dfrac{1}{2} \sqrt{-\dfrac{g^{11}}{g_{00}}} \dfrac{dg_{00}}{dr} \Biggr) 	\nonumber\\
=& \frac{ \left[ (r_H-M)-\frac{\Lambda~ r_H}{3} \left(2 r_H^{2}+a^2 \right) \right]}{   \left(r^{2}_{H}+a^2\right) \left(1+\frac{\Lambda a^2}{3} \right)}.
 \end{align}
 The original Hawking temperature of the KNdS black hole is obtained from the relation $T_o=\frac{\kappa}{2 \pi}$ \cite{christina1} as
 \begin{align}
 T_o=\frac{ \left[ (r_H-M)-\frac{\Lambda~ r_H}{3} \left(2 r_H^{2}+a^2 \right) \right]}{2\pi   \left(r^{2}_{H}+a^2\right) \left(1+\frac{\Lambda a^2}{3} \right)}.
 \end{align}

By using the expression of $M$ from eq. \eqref{eqn 6}, we derive the heat capacity of the black hole by using the relation $C_o=\frac{\partial M}{\partial T_o}$ \cite{priyo1, arun1} as

\begin{align}
C_o=&\left(\dfrac{\partial M}{\partial r_h} \right) \left( \dfrac{\partial r_h}{\partial T_o} \right) \nonumber\\
=& \dfrac{2 \pi \left(a^2+r^2\right)^2 \left(3+a^2 \Lambda\right) \left[ a^2 \left( 3+r^2 \Lambda \right)+3 \left(Q^2-r^2+r^4 \Lambda \right)\right]}{3 \left[ a^4 \left(r^2 \Lambda -3\right) +3 r^2 \left( r^2+r^4 \Lambda-3 Q^2\right)+a^2 \left( 8 r^4 \Lambda -3 Q^2-12 r^2 \right) 			\right]}.  
\end{align}

Applying WKB approximation, $\Psi_\mu$  given in eq. \eqref{eqn 3} is taken as
\begin{align} \label{eqn 14}
\Psi_\mu= c_\mu (t,r,\theta) \exp\left[\frac{i}{\hbar} \mathcal{S} (t,r,\theta)	\right],	
\end{align}
where $\mathcal{S}$ is defined as
\begin{align} \label{eqn 15}
\mathcal{S}(t,r,\theta)=\mathcal{S}_0(t,r,\theta)+\hbar ~ \mathcal{S}_1(t,r,\theta)+ \hbar^2 ~ \mathcal{S}_2 (t,r,\theta) + \cdots.
\end{align}

By substituting eqs. \eqref{eqn 14}, \eqref{eqn 15} and \eqref{eqn 11} in eq. \eqref{eqn 3} and keeping only the lowest order in $\hbar$, we obtain the equations for the coefficients $c_\mu$ as

%  g^{00}=		-\dfrac{\Sigma^2  \left(r^{2}_{H}+a^2\right)^2}{(r-r_H) \Delta,_r(r_H)  ~\rho^2(r_H)}
%
% 	 g^{11}=	\dfrac{(r-r_H) \Delta,_r(r_H)}{\rho^2(r_H)}
%
% g^{22}=		\dfrac{\Delta_\theta}{\rho^2(r_H)}

\begin{align}\label{eqn 16}
 & \dfrac{(r-r_H ) \Delta,_r(r_H)}{\rho^2(r_H)} \left[ c_0~ \mathcal{B}_{1}^2 \left(\partial_r \mathcal{S}_0\right)^2- c_1~ \mathcal{B}_0 \mathcal{B}_1 \left(\partial_r \mathcal{S}_0	\right)	(\partial_t \mathcal{S}_0 +e A_0)	\right] \nonumber \\
& ~~~~~~ + \dfrac{\Delta_\theta}{\rho^2(r_H)} \left[	c_0 ~ \mathcal{B}_{2}^2 (\partial_\theta \mathcal{S}_0)^2	 -c_2 \mathcal{B}_0 \mathcal{B}_2 (\partial_\theta \mathcal{S}_0)(\partial_t \mathcal{S}_0 +e A_0)		\right] +m^2 c_0~=~0, 
\end{align}
\begin{align}
& -\dfrac{\Sigma^2  \left(r^{2}_{H}+a^2\right)^2}{(r-r_H) \Delta,_r(r_H)  ~\rho^2(r_H)} \left[ c_1~ \mathcal{B}_{0}^2 (\partial_t \mathcal{S}_0 +e A_0)^2- c_0~ \mathcal{B}_0 \mathcal{B}_1 \left(\partial_r \mathcal{S}_0	\right)	(\partial_t \mathcal{S}_0 +e A_0)	\right] \nonumber \\
& ~~~~~~ + \dfrac{\Delta_\theta}{\rho^2(r_H)} \left[	c_1 ~ \mathcal{B}_{2}^2 (\partial_\theta \mathcal{S}_0)^2	 -c_2 \mathcal{B}_1 \mathcal{B}_2 (\partial_r  \mathcal{S}_0) (\partial_\theta \mathcal{S}_0)		\right] +m^2 c_1~=~0, 
\end{align}
\begin{align}\label{eqn 18}
& -\dfrac{\Sigma^2  \left(r^{2}_{H}+a^2\right)^2}{(r-r_H) \Delta,_r(r_H)  ~\rho^2(r_H)} \left[ c_2~ \mathcal{B}_{0}^2 (\partial_t \mathcal{S}_0 +e A_0)^2- c_0~ \mathcal{B}_0 \mathcal{B}_2 \left(\partial_\theta \mathcal{S}_0	\right)	(\partial_t \mathcal{S}_0 +e A_0)	\right] \nonumber \\
& ~~~~~~ + \dfrac{(r-r_H) \Delta,_r(r_H)}{\rho^2(r_H)} \left[	c_2 ~ \mathcal{B}_{1}^2 (\partial_r \mathcal{S}_0)^2	 -c_1 \mathcal{B}_1 \mathcal{B}_2 (\partial_r  \mathcal{S}_0) (\partial_\theta \mathcal{S}_0)		\right] +m^2 c_2~=~0,
\end{align}
where  the $\mathcal{B}_\mu$'s are defined as
\begin{align} \label{eqn 19}
& \mathcal{B}_0=1+\dfrac{\beta  ~\Sigma^2  \left(r^{2}_{H}+a^2\right)^2 (\partial_t \mathcal{S}_0 + e A_0)^2}{(r-r_H) \Delta,_r(r_H)  ~\rho^2(r_H)}, \nonumber \\
& \mathcal{B}_1=1+ \dfrac{ \beta ~(r-r_H) \Delta,_r(r_H) (\partial_r \mathcal{S}_0)^2}{\rho^2(r_H)},~~~ \mathcal{B}_2=1+\dfrac{\beta \Delta_\theta (\partial_\theta \mathcal{S}_0)^2}{\rho^2(r_H)}.
\end{align}

Considering the symmetries of the  metric \eqref{eqn 11}, we carry on the separation of variables as follows
\begin{align} \label{eqn 20}
\mathcal{S}_0=- (E-j \Omega) ~t+ R (r)+ \Theta (\theta)+U,
\end{align}
where $E$ is the energy of the emitted vector particles, $j$ is the angular momentum and $U$ is a complex constant. On inserting eq. \eqref{eqn 20} in eqs. \eqref{eqn 16}-\eqref{eqn 18}, we get a matrix equation as
\begin{align} \label{eqn 21}
F(c_0,c_1,c_2)^T~=~0,
\end{align}
where $F$ is a $3 \times 3$ matrix and all the entries are as follows
%  g^{00}=		-\dfrac{\Sigma^2  \left(r^{2}_{H}+a^2\right)^2}{(r-r_H) \Delta,_r(r_H)  ~\rho^2(r_H)}
%
% 	 g^{11}=	\dfrac{(r-r_H) \Delta,_r(r_H)}{\rho^2(r_H)}
%
% g^{22}=		\dfrac{\Delta_\theta}{\rho^2(r_H)}
\begin{align}
 F_{11}=&\dfrac{(r-r_H) \Delta,_r(r_H)}{\rho^2(r_H)} \mathcal{B}_1^2 \mathcal{R}'^{2}+ \dfrac{\Delta_\theta}{\rho^2(r_H)} \mathcal{B}_{2}^2 J_{\theta}^2+m^2, \nonumber\\
 F_{12}=&-\dfrac{(r-r_H) \Delta,_r(r_H)}{\rho^2(r_H)}(-E+j \Omega+e A_0) \mathcal{B}_0 \mathcal{B}_1 \mathcal{R}', \nonumber\\
 F_{13}=&- \dfrac{\Delta_\theta}{\rho^2(r_H)} (-E+j \Omega+e A_0) \mathcal{B}_0 \mathcal{B}_2 J_{\theta}, \nonumber\\
 F_{21}= &\dfrac{\Sigma^2  \left(r^{2}_{H}+a^2\right)^2}{(r-r_H) \Delta,_r(r_H)  ~\rho^2(r_H)}  (-E+j \Omega+e A_0) \mathcal{B}_0 \mathcal{B}_1 \mathcal{R}', \nonumber\\
 F_{22}=&-\dfrac{\Sigma^2  \left(r^{2}_{H}+a^2\right)^2}{(r-r_H) \Delta,_r(r_H)  ~\rho^2(r_H)} (-E+j \Omega+e A_0)^2 \mathcal{B}_0^2+ \dfrac{\Delta_\theta}{\rho^2(r_H)} \mathcal{B}_{2}^{2} J_{\theta}^2+m^2, \nonumber\\
 F_{23}=& -\dfrac{\Delta_\theta}{\rho^2(r_H)} \mathcal{B}_1 \mathcal{B}_2 J_{\theta} \mathcal{R}', \nonumber\\
 F_{31}=&\dfrac{\Sigma^2  \left(r^{2}_{H}+a^2\right)^2}{(r-r_H) \Delta,_r(r_H)  ~\rho^2(r_H)} (-E+j \Omega+e A_0) \mathcal{B}_0 \mathcal{B}_2 J_{\theta} , \nonumber\\
 F_{32}=&-\dfrac{(r-r_H) \Delta,_r(r_H)}{\rho^2(r_H)} \mathcal{B}_1 \mathcal{B}_2 J_{\theta} \mathcal{R}', \nonumber\\
 F_{33}=& -\dfrac{\Sigma^2  \left(r^{2}_{H}+a^2\right)^2}{(r-r_H) \Delta,_r(r_H)  ~\rho^2(r_H)} (-E+j \Omega+e A_0)^2 \mathcal{B}_2+ \dfrac{(r-r_H) \Delta,_r(r_H)}{\rho^2(r_H)} \mathcal{B}_{1}^2 \mathcal{R}'^2 \nonumber\\ & +m^2,  
\end{align}
where $\mathcal{R}'=\partial_{r} \mathcal{R}$ and $J_{\theta}=\partial_{\theta} \Theta$.

To find a non-trivial solution of eq. \eqref{eqn 21}, we put the determinant of the matrix $F$ equals zero, which in turn gives
\begin{align} \label{eqn 23}
\mathcal{O}(\beta^4) & (\partial_r \mathcal{R})^{12} + \mathcal{O} (\beta^3) (\partial_r \mathcal{R})^{10}+ \mathcal{O}(\beta^2) (\partial_r \mathcal{R})^8+ \left[ H_6+ \mathcal{O}(\beta^2)		\right](\partial_r \mathcal{R})^6 \nonumber\\
&+ \left[ H_4+ \mathcal{O}(\beta^2)		\right](\partial_r \mathcal{R})^4+\left[ H_2+ \mathcal{O}(\beta^2)		\right](\partial_r \mathcal{R})^2 +H_0 + \mathcal{O}(\beta^2)	~=~0,
\end{align}

%  g^{00}=		-\dfrac{\Sigma^2  \left(r^{2}_{H}+a^2\right)^2}{(r-r_H) \Delta,_r(r_H)  ~\rho^2(r_H)}
%
% 	 g^{11}=	\dfrac{(r-r_H) \Delta,_r(r_H)}{\rho^2(r_H)}
%
% g^{22}=		\dfrac{\Delta_\theta}{\rho^2(r_H)}

where
\begin{align*}
 H_6= & 4 \beta \left[\dfrac{(r-r_H) \Delta,_r(r_H)}{\rho^2(r_H)}	\right]^3, \nonumber\\ 
 H_4=& \dfrac{(r-r_H)^2 \Delta^{2}_{,_r}(r_H)}{\rho^4(r_H)}   \left[	1+ 4 \beta  \left\lbrace   \dfrac{ -  (-E+j \Omega+e A_0)^2 ~\Sigma^2  \left(r^{2}_{H}+a^2\right)^2}{(r-r_H) \Delta,_r(r_H)  ~\rho^2(r_H)}	+ \dfrac{\Delta_\theta J_{\theta}^2}{\rho^2(r_H)}  + m^2 \right\rbrace
\right]	, 
\end{align*}
\begin{align*}
 H_2= & \dfrac{2(r-r_H) \Delta,_r(r_H)}{\rho^2(r_H)}	 \left[ \dfrac{ -  (-E+j \Omega+e A_0)^2 ~\Sigma^2  \left(r^{2}_{H}+a^2\right)^2}{(r-r_H) \Delta,_r(r_H)  ~\rho^2(r_H)}+	 	\dfrac{\Delta_\theta J_{\theta}^2}{\rho^2(r_H)}+m^2 \right. \nonumber \\
 & \left.			 
 -2 \beta \left\lbrace \dfrac{   (-E+j \Omega+e A_0)^4 ~\Sigma^4  \left(r^{2}_{H}+a^2\right)^4}{(r-r_H)^2 \Delta^{2},_r(r_H)  ~\rho^4(r_H)}-	 	\dfrac{\Delta_\theta^2 J_{\theta}^4}{\rho^2(r_H)}  \right\rbrace \right], \nonumber\\
H_0=	& \left\lbrace   \dfrac{ -  (-E+j \Omega+e A_0)^2 ~\Sigma^2  \left(r^{2}_{H}+a^2\right)^2}{(r-r_H) \Delta,_r(r_H)  ~\rho^2(r_H)}	+ \dfrac{\Delta_\theta J_{\theta}^2}{\rho^2(r_H)}  + m^2 \right\rbrace  \left[ 	 \dfrac{\Delta_\theta J_{\theta}^2}{\rho^2(r_H)}  + m^2 	 \right. \nonumber \\
 & \left. +  \dfrac{(-E+j \Omega+e A_0)^2 ~\Sigma^2  \left(r^{2}_{H}+a^2\right)^2}{(r-r_H) \Delta,_r(r_H)  ~\rho^2(r_H)}			 -4 \beta \left\lbrace  \dfrac{	  (-E+j \Omega+e A_0)^4 ~\Sigma^4  \left(r^{2}_{H}+a^2\right)^4}{(r-r_H)^2 \Delta^{2},_r(r_H)  ~\rho^4(r_H)}\right.\right. \nonumber \\
 & \left. \left.	- \dfrac{\Delta^{2}_{\theta} J_{\theta}^4}{\rho^4(r_H)} 		\right\rbrace	\right].
\end{align*}

By neglecting the  higher order term of $\beta$, eq. \eqref{eqn 23} becomes

\begin{align}
& \dfrac{4 \beta(r-r_H)^2 \Delta^{2},_r(r_H) ~ (\partial_r \mathcal{R})^{4}}{\rho^4(r_H)} + \dfrac{(r-r_H) \Delta,_r(r_H) ~ (\partial_r \mathcal{R})^{2}}{\rho^2(r_H)}+\dfrac{\Delta_\theta J^{2}_{\theta}}{\rho^2(r_H)}  \nonumber\\
&  + 4 \beta \left\lbrace  \dfrac{\Delta^{2}_\theta J^{4}_{\theta}}{\rho^4(r_H)}- \dfrac{(-E+j \Omega+e A_0)^4 {}{}{}{}{}{} \Sigma^4  \left(r^{2}_{H}+a^2\right)^4}{(r-r_H)^2 \Delta^{2},_r(r_H)  ~\rho^4(r_H)} 	\right\rbrace  +m^2 \nonumber\\ & -\dfrac{(-E +j \Omega+e A_0)^2 \Sigma^2  \left(r^{2}_{H}+a^2\right)^2}{(r-r_H) \Delta,_r(r_H)  ~\rho^2(r_H)} =0.
\end{align}

%\begin{align}
%& g= -\dfrac{\rho^6 (r_H)}{\Sigma^2 \left(r_{H}^2+a^2\right)^2 \Delta_{\theta}}, \nonumber \\
%& g^{00}=-\dfrac{\Sigma^2  \left(r^{2}_{H}+a^2\right)^2}{(r-r_H) \Delta,_r(r_H)  ~\rho^2(r_H)}, \nonumber\\
%& g^{11}=\dfrac{(r-r_H) \Delta,_r(r_H)}{\rho^2(r_H)},~~ g^{22}=\dfrac{\Delta_\theta}{\rho^2(r_H)}, \nonumber\\
%& g^{01}=g^{02}=g^{10}=g^{12}=g^{20}=g^{21}=0.
%\end{align}

We obtain the solution to the derivative of the radial action by neglecting the higher power of $\beta$ as

%By neglecting the higher order term of $\beta$ and solving Eq. \eqref{eqn 23}, we get the solution to the derivative of the radial action
\begin{align} \label{eqn 25}
\partial_r \mathcal{R}=\pm \sqrt{-\dfrac{\left(J^2_{\theta} \Delta_{\theta} +m^2 \rho^2		\right)}{(r-r_{H})~\Delta,_r(r_H)}	+	\dfrac{(-E +j \Omega+e A_0)^2 \left(r^{2}_{H}+a^2\right)^2~ \Sigma^2}{(r-r_{H})^2~\left(\Delta,_r(r_H)\right)^2}}  \left(1+\beta~\dfrac{\chi_1}{\chi_2}\right),
\end{align}
where
\begin{align}
\chi_1= & 2 \Delta,_r(r_H) (r-r_H)			
  \left(	 2 J^{4}_\theta \Delta^{2}_\theta +2 J^{2}_\theta m^2 \Delta_\theta  \rho^2 +  m^4 \rho^4		\right)	 	\nonumber\\
  & -4 \Sigma^2 (-E +j \Omega+e A_0)^2  \left(r^{2}_{H}+a^2\right)^2 \left(  J^{2}_\theta  \Delta_{\theta} +m^2 \rho^2 \right), \nonumber\\
\chi_2= & \rho^2 \left\lbrace	  (r-r_H) \Delta,_r(r_H) \left(	J^2_{\theta} \Delta_\theta +m^2 \rho^2 \right)	- \Sigma^2 (-E +j \Omega+e A_0)^2 \left(r^{2}_{H}+a^2\right)^2		\right\rbrace.
\end{align}

The solution of the radial action is obtained by integrating eq. \eqref{eqn 25} around the pole $r=r_H$. The imaginary part of the radial action gives the particle's rate of tunneling as 
\begin{align} \label{eqn 27}
Im~\mathcal{R}_\pm =& \pm Im ~ \int dr \sqrt{-\dfrac{\left(J^2_{\theta} \Delta_{\theta} +m^2 \rho^2		\right)}{(r-r_{H})~\Delta,_r(r_H)}	+	\dfrac{(-E +j \Omega+e A_0)^2 ~\left(r^{2}_{H}+a^2\right)^2~ \Sigma^2}{(r-r_{H})^2~\left(\Delta,_r(r_H)\right)^2}}~ \nonumber\\
& \times \left(1+\beta~\dfrac{\chi_1}{\chi_2}\right)
\nonumber \\ =&\pm \dfrac{\pi  (E -j \Omega-e A_0) \left(r^{2}_{H}+a^2\right) \left(1+\frac{\Lambda a^2}{3} \right)}{2 \left[ (r_H-M)-\frac{\Lambda~ r_H}{3} \left(2 r_H^{2}+a^2 \right) \right]} \left( 1+ \beta ~ \Pi	\right),
\end{align}
%where $\mathfrak{X}=\dfrac{m(1+2m)}{m-1}+\dfrac{J^2_{\theta} \Delta_\theta (3+2m)}{(m-1) \left(r^{2}_H + a^2 \cos^2\theta \right)}$.
where $\Pi=4m^2+\frac{4 J_{\theta}^{2} \Delta_{\theta}}{\rho^2}$. $\mathcal{R}_+$ and $\mathcal{R}_-$ stand for the radial action of the outgoing and ingoing particles respectively. Based on WKB approximation, the probabilities of the vector boson particles tunneling across the event horizon $r_H$ are
\begin{align}
& \mathcal{P}_{outgoing}= \exp \biggl[ -2 \biggl\{Im (\mathcal{R}_+) + Im(U) \biggr\} 		\biggr] \nonumber\\
& \text{and} \nonumber\\
& \mathcal{P}_{ingoing}=\exp \biggl[ -2 \biggl\{Im (\mathcal{R}_-) + Im(U) \biggr\} 		\biggr].
\end{align}
According to the semiclassical WKB approximation, the ingoing vector boson particles have a $100\%$ chance of entering the black hole \cite{a6}. It shows that $Im(\mathcal{R}_+)=-Im(\mathcal{R}_-)$. Thus the tunneling rate of the vector boson particles is given by
\begin{align}
\Gamma=\dfrac{\mathcal{P}_{outgoing}}{\mathcal{P}_{ingoing}}=  \exp & \left[ -4~ Im ~\mathcal{R}_+	\right] 	\nonumber\\
= \exp &\left[ 	\dfrac{-2\pi  (E -j \Omega-e A_0) \left(r^{2}_{H}+a^2\right) \left(1+\frac{\Lambda a^2}{3} \right)}{ \left[ (r_H-M)-\frac{\Lambda~ r_H}{3} \left(2 r_H^{2}+a^2 \right) \right]} 	(1+\beta ~\Pi)	\right].
\end{align}

The Hawking temperature of the KNdS  black hole under the quantum gravity effect is obtained as
\begin{align} \label{eq 34}
T_d=& \dfrac{ \left[ (r_H-M)-\frac{\Lambda~ r_H}{3} \left(2 r_H^{2}+a^2 \right) \right]}{2\pi   \left(r^{2}_{H}+a^2\right) \left(1+\frac{\Lambda a^2}{3} \right)} \left(1- \beta ~ \Pi \right) \nonumber\\
=& T_o ~\left(1- \beta ~ \Pi \right),
\end{align}
where $T_o=\frac{ \left[ (r_H-M)-\frac{\Lambda~ r_H}{3} \left(2 r_H^{2}+a^2 \right) \right]}{2\pi   \left(r^{2}_{H}+a^2\right) \left(1+\frac{\Lambda a^2}{3} \right)}$ is the original Hawking temperature. It is obvious from the point $\Pi>0$ that the Hawking temperature is modified due to the quantum gravity effects. The modifed Hawking temperature is lower than the original Hawking temperature due to the presence of $\beta>0$. Further, the modified Hawking temperature relies on the mass and angular momentum of the emitted vector boson particles. In the absence of the quantum gravity effects i.e. $\beta=0$, the original Hawking temperature of KNdS black hole is recovered.

The modified heat capacity is calculated as 
\begin{align} \label{eqn 35}
C_H=& \dfrac{2 \pi \left(a^2+r^2\right)^2 \left(3+a^2 \Lambda\right) \left[ a^2 \left( 3+r^2 \Lambda \right)+3 \left(Q^2-r^2+r^4 \Lambda \right)\right]}{3 \left[ a^4 \left(r^2 \Lambda -3\right) +3 r^2 \left( r^2+r^4 \Lambda-3 Q^2\right)+a^2 \left( 8 r^4 \Lambda -3 Q^2-12 r^2 \right) 			\right]} \nonumber \\
& \times \left( 1+ \beta ~\Pi \right) \nonumber \\
=& C_o \left( 1+ \beta ~\Pi \right).
\end{align}
From eq. \eqref{eqn 35}, it is observed that the modified heat capacity reduced to the original heat capacity ($C_o$) when $\beta=0$. The modified heat capacity of KNdS black hole increases due to the quantum gravity effects.

Now using another coordinate transformation $\phi=\varphi- \Omega t$ \cite{chen3} where
\begin{align}
\Omega=\dfrac{a~  [(r^2+a^2) \Delta_{\theta}-\Delta]}{(r^2+a^2)^2~ \Delta_{\theta}-\Delta a^2\sin^2\theta },
\end{align}

%Now we perform another frame dragging transformation  in Eq. \eqref{eqn 4}. Let $\phi=\varphi- \Omega t$ and  $\Omega=\frac{-g_{00}}{g_{03}}$, then Eq. \eqref{eqn  4} reduced to
then the line element \eqref{eqn 4} reduces to 
\begin{align} \label{eqn 36}
ds^2=& -\dfrac{ \Delta \Delta_{\theta} \rho^2}{ \Sigma^2 \left[ \Delta_\theta(r^2+a^2)^2-\Delta a^2\sin^2\theta \right]} dt^2 +\dfrac{\rho^2}{\Delta}dr^2+ \dfrac{\rho^2}{\Delta_\theta} d\theta^2 \nonumber \\
& +\dfrac{\left[ \Delta_\theta(r^2+a^2)^2-\Delta a^2\sin^2\theta \right] \sin^2\theta}{\rho^2 \Sigma^2} d\phi^2.
\end{align}
The corresponding non-zero electromagnetic potentials are given by
\begin{align} \label{eqn 37}
A_0=\dfrac{(r^2+a^2)\Delta_\theta 	~ Q ~r}{\Sigma \left[\Delta_\theta (r^2+a^2)^2-\Delta a^2 \sin^2\theta \right]},\qquad A_3=-\dfrac{Q r a \sin^2 \theta}{(r^2+a^2 \cos^2 \theta) \Sigma}.
\end{align}
Using  eq. \eqref{eqn 10} in eq. \eqref{eqn 36}, we get
%On applying the approximation eq. \eqref{eqn 10}, the line element eq. \eqref{eqn 36} gives
\begin{align}\label{eqn 38}
ds^2=& -\dfrac{(r-r_H) \Delta,_r(r_H)  ~\rho^2(r_H)}{\Sigma^2  \left(r^{2}_{H}+a^2\right)^2} dt^2+\dfrac{\rho^2(r_H)}{(r-r_H) \Delta,_r(r_H)} dr^2+\dfrac{\rho^2(r_H)}{\Delta_\theta}d\theta^2 \nonumber\\
& + \dfrac{\Delta_{\theta} \left(r^{2}_{H}+a^2\right)^2 \sin^2\theta }{\rho^2(r_H) \Sigma^2} d\phi^2.
\end{align}

%The nonvanishing electric potentials near the event horizon are given as
%\begin{align} \label{eqn 39}
%A_0=\dfrac{Q~ r_{H}}{\Sigma  \left(r^{2}_{H}+a^2\right) },\qquad A_3=\dfrac{-Q r a \sin^2 \theta}{(r^2+a^2 \cos^2 \theta) \Sigma}.
%\end{align}

The metric determinant and the nonzero contravariant components of the line element given above are as follows
\begin{align}
& g=\dfrac{\rho^4 \sin^2 \theta}{\Sigma^4}, \quad  g^{00}=-\dfrac{\Sigma^2  \left(r^{2}_{H}+a^2\right)^2}{(r-r_H) \Delta,_r(r_H)  ~\rho^2(r_H)}, \nonumber\\
& g^{11}=\dfrac{(r-r_H) \Delta,_r(r_H)}{\rho^2(r_H)}, \quad g^{22}=\dfrac{\Delta_\theta}{\rho^2(r_H)}, 
\nonumber\\
& g^{33}=\dfrac{\rho^2(r_H) \Sigma^2}{\Delta_{\theta} \left(r^{2}_{H}+a^2\right)^2 \sin^2\theta}.
\end{align}

According to WKB approximation, $\Psi_{\mu}$  from eq. \eqref{eqn 3} can be written as
\begin{align} \label{eqn 41}
\Psi_\mu= c_\mu (t,r,\theta,\phi) \exp\left[\frac{i}{\hbar} \mathcal{S} (t,r,\theta,\phi)	\right],	
\end{align}
where $\mathcal{S}$ is defined as
\begin{align} \label{eqn 42}
\mathcal{S}(t,r,\theta)=\mathcal{S}_0(t,r,\theta,\phi)+\hbar ~ \mathcal{S}_1(t,r,\theta,\phi)+ \hbar^2 ~ \mathcal{S}_2 (t,r,\theta,\phi) + \cdots.
\end{align}

By substituting eqs. \eqref{eqn 41} and \eqref{eqn 42} in eq. \eqref{eqn 38}, and keeping only the lowest order in $\hbar$, we derive the equations for the coefficients $c_\mu$ as

\begin{align}\label{eqn 43}
 & \dfrac{(r-r_H ) \Delta,_r(r_H)}{\rho^2(r_H)} \left[ c_0~ \mathcal{H}_{1}^2 \left(\partial_r \mathcal{S}_0\right)^2- c_1~ \mathcal{H}_0 \mathcal{H}_1 \left(\partial_r \mathcal{S}_0	\right)	(\partial_t \mathcal{S}_0 +e A_0)	\right]   + \dfrac{\Delta_\theta}{\rho^2(r_H)} 
 \nonumber \\
& \left[	c_0 ~ \mathcal{H}_{2}^2~ (\partial_\theta \mathcal{S}_0)^2	 -c_2 \mathcal{H}_0 \mathcal{H}_2  ~ (\partial_\theta \mathcal{H})(\partial_t \mathcal{S}_0 +e A_0)		\right] + \dfrac{\rho^2(r_H) ~ \Sigma^2}{\Delta_{\theta} \left(r^{2}_{H}+a^2\right)^2 \sin^2\theta}\nonumber\\
& \left[	c_0 	\mathcal{H}_{3}^2 ~(\partial_\phi \mathcal{S}_0 +e A_3)^2	- c_3  ~ \mathcal{H}_0 \mathcal{H}_3 \left(\partial_t \mathcal{S}_0 +e A_0	\right)	(\partial_\phi \mathcal{S}_0 +e A_3) \right]  +m^2 c_0~=~0,
\end{align}
\begin{align}\label{eqn 44}
& -\dfrac{\Sigma^2  \left(r^{2}_{H}+a^2\right)^2}{(r-r_H) \Delta,_r(r_H)  ~\rho^2(r_H)} \left[ c_1~ \mathcal{H}_{0}^2 (\partial_t \mathcal{S}_0 +e A_0)^2- c_0~ \mathcal{H}_0 \mathcal{H}_1 \left(\partial_r \mathcal{S}_0	\right)	(\partial_t \mathcal{S}_0 +e A_0)	\right] \nonumber \\
& + \dfrac{\Delta_\theta}{\rho^2(r_H)} \left[	c_1 ~ \mathcal{H}_{2}^2 (\partial_\theta \mathcal{S}_0)^2	 -c_2 \mathcal{H}_1 \mathcal{H}_2 (\partial_r  \mathcal{S}_0) (\partial_\theta \mathcal{S}_0)		\right]  +\dfrac{\rho^2(r_H) \Sigma^2}{\Delta_{\theta} \left(r^{2}_{H}+a^2\right)^2 \sin^2\theta} \nonumber\\
& \left[ c_1~ \mathcal{H}_{3}^2 (\partial_\phi \mathcal{S}_0 +e A_3)^2- c_3~ \mathcal{H}_1 \mathcal{H}_3 \left(\partial_r \mathcal{S}_0	\right)	(\partial_\phi \mathcal{S}_0 +e A_3)	\right]+m^2 c_1~=~0, 
\end{align}
\begin{align}\label{eqn 45}
& -\dfrac{\Sigma^2  \left(r^{2}_{H}+a^2\right)^2}{(r-r_H) \Delta,_r(r_H)  ~\rho^2(r_H)} \left[ c_2~ \mathcal{H}_{0}^2 (\partial_t \mathcal{S}_0 +e A_0)^2- c_0~ \mathcal{H}_0 \mathcal{H}_2 \left(\partial_\theta \mathcal{S}_0	\right)	(\partial_t \mathcal{S}_0 +e A_0)	\right] \nonumber \\
&  + \dfrac{(r-r_H) \Delta,_r(r_H)}{\rho^2(r_H)} \left[	c_2 ~ \mathcal{H}_{1}^2 (\partial_r \mathcal{S}_0)^2	 -c_1 \mathcal{H}_1 \mathcal{H}_2 (\partial_r  \mathcal{S}_0) (\partial_\theta \mathcal{S}_0)		\right] +   \dfrac{\rho^2(r_H) \Sigma^2}{\Delta_{\theta} \left(r^{2}_{H}+a^2\right)^2 \sin^2\theta} \nonumber\\
&  \left[ c_2~ \mathcal{H}_{3}^2 (\partial_\phi \mathcal{S}_0 +e A_3)^2- c_3~ \mathcal{H}_2 \mathcal{H}_3 \left(\partial_\theta \mathcal{S}_0	\right)	(\partial_\phi \mathcal{S}_0 +e A_3)	\right]   +m^2 c_2~=~0,
\end{align}

%\begin{align}
% &g^{00}=   - \dfrac{\Sigma^2  \left(r^{2}_{H}+a^2\right)^2}{(r-r_H) \Delta,_r(r_H)  ~\rho^2(r_H)}\nonumber\\ 
%& g^{11}=    \dfrac{(r-r_H) \Delta,_r(r_H)}{\rho^2(r_H)} 
%\nonumber\\ 
% g^{22}=      \dfrac{\Delta_\theta}{\rho^2(r_H)}
%\nonumber\\ 
%& g^{33}=      \dfrac{\rho^2(r_H) \Sigma^2}{\Delta_{\theta} \left(r^{2}_{H}+a^2\right)^2 \sin^2\theta}
%\end{align}

\begin{align}\label{eqn 46}
& -\dfrac{\Sigma^2  \left(r^{2}_{H}+a^2\right)^2}{(r-r_H) \Delta,_r(r_H)  ~\rho^2(r_H)} \left[ c_3~ \mathcal{H}_{0}^2 (\partial_t \mathcal{S}_0 +e A_0)^2- c_0~ \mathcal{H}_0 \mathcal{H}_3 	(\partial_t \mathcal{S}_0 +e A_0)  \right. \nonumber \\
& \left. (\partial_\phi \mathcal{S}_0 +e A_3)	\right]  + \dfrac{(r-r_H) \Delta,_r(r_H)}{\rho^2(r_H)} \left[	c_3 ~ \mathcal{H}_{1}^2 (\partial_r \mathcal{S}_0)^2	 -c_1 \mathcal{H}_1 \mathcal{H}_3 (\partial_r  \mathcal{S}_0) (\partial_\phi \mathcal{S}_0 +e A_3)		\right] \nonumber \\
&  + \dfrac{\Delta_\theta}{\rho^2(r_H)} 
  \left[	c_3 ~ \mathcal{H}_{2}^2~ (\partial_\theta \mathcal{S}_0)^2	 -c_2 \mathcal{H}_2 \mathcal{H}_3  ~ (\partial_\theta \mathcal{S}_0)(\partial_\phi \mathcal{S}_0 +e A_3)		\right]   +m^2 c_3~=~0,
\end{align}

%\begin{align}
% &g^{00}=   - \dfrac{\Sigma^2  \left(r^{2}_{H}+a^2\right)^2}{(r-r_H) \Delta,_r(r_H)  ~\rho^2(r_H)}\nonumber\\ 
%& g^{11}=    \dfrac{(r-r_H) \Delta,_r(r_H)}{\rho^2(r_H)} 
%\nonumber\\ 
% g^{22}=      \dfrac{\Delta_\theta}{\rho^2(r_H)}
%\nonumber\\ 
%& g^{33}=      \dfrac{\rho^2(r_H) \Sigma^2}{\Delta_{\theta} \left(r^{2}_{H}+a^2\right)^2 \sin^2\theta}
%\end{align}
where  the $\mathcal{H}_\mu$'s are defined as
\begin{align} \label{eqn 47}
& \mathcal{H}_0=1+\beta  \dfrac{ ~\Sigma^2  \left(r^{2}_{H}+a^2\right)^2 (\partial_t \mathcal{S}_0 + e A_0)^2}{(r-r_H) \Delta,_r(r_H)  ~\rho^2(r_H)}, ~~ \mathcal{H}_1=1+ \beta \dfrac{  (r-r_H) \Delta,_r(r_H) (\partial_r \mathcal{S}_0)^2}{\rho^2(r_H)},\nonumber \\
&  \mathcal{H}_2=1+\beta~ \dfrac{ \Delta_\theta~ (\partial_\theta \mathcal{S}_0)^2}{\rho^2(r_H)}, ~~ \quad\mathcal{H}_3=1+ \beta~ \dfrac{\rho^2(r_H) \Sigma^2~ (\partial_\phi \mathcal{S}_0 + e A_3)^2}{\Delta_{\theta} \left(r^{2}_{H}+a^2\right)^2 \sin^2\theta}.
\end{align}

Considering the symmetry of the black hole, the corresponding action of the vector boson particles can be written as
\begin{align} \label{eqn 49}
\mathcal{S}_0=- (E-j \Omega) ~t+ \mathcal{W} (r)+ \Theta (\theta, \phi)+V,
\end{align}
where $E$, $j$ and $V$ are the energy of the emitted vector particles, angular momentum and complex constant respectively. On inserting eq. \eqref{eqn 49} in eqs. \eqref{eqn 43}-\eqref{eqn 46}, a matrix equation is obtained as 
\begin{align} \label{eqn 50}
\mathcal{F} \left(c_0, c_1, c_2, c_3\right)^T=0,
\end{align}
where $\mathcal{F}$ is a $4 \times 4$ matrix, whose components of the matrix are as follows

%\begin{align}
% &g^{00}=   - \dfrac{\Sigma^2  \left(r^{2}_{H}+a^2\right)^2}{(r-r_H) \Delta,_r(r_H)  ~\rho^2(r_H)}
%\nonumber\\ 
%& g^{11}=    \dfrac{(r-r_H) \Delta,_r(r_H)}{\rho^2(r_H)} 
%\nonumber\\ 
% g^{22}=      \dfrac{\Delta_\theta}{\rho^2(r_H)}
%\nonumber\\ 
%& g^{33}=      \dfrac{\rho^2(r_H) \Sigma^2}{\Delta_{\theta} \left(r^{2}_{H}+a^2\right)^2 \sin^2\theta}
%\end{align}

\begin{align} \label{eqn 51}
 \mathcal{F}_{11}=&\dfrac{(r-r_H) \Delta,_r(r_H)}{\rho^2(r_H)} \mathcal{H}_1^2~ \mathcal{W}'^{2}+ \dfrac{\Delta_\theta}{\rho^2(r_H)} \mathcal{H}_{2}^2~ J_{\theta}^2+   \dfrac{\rho^2(r_H) \Sigma^2}{\Delta_{\theta} \left(r^{2}_{H}+a^2\right)^2 \sin^2\theta}  \nonumber\\ &  \mathcal{H}_{3}^2~ \left(J_\phi +e A_3\right)         +m^2, \nonumber\\
 \mathcal{F}_{12}=&-\dfrac{(r-r_H) \Delta,_r(r_H)}{\rho^2(r_H)}(-E+j \Omega+e A_0) \mathcal{H}_0 \mathcal{H}_1 \mathcal{W}', \nonumber\\
 \mathcal{F}_{13}=&- \dfrac{\Delta_\theta}{\rho^2(r_H)} (-E+j \Omega+e A_0) \mathcal{H}_0 \mathcal{H}_2~ J_{\theta}, \nonumber\\
  \mathcal{F}_{14}=&-    \dfrac{\rho^2(r_H) ~ \Sigma^2}{\Delta_{\theta} \left(r^{2}_{H}+a^2\right)^2 \sin^2\theta}         (-E+j \Omega+e A_0) \left( J_{\phi}+e A_3\right) ~\mathcal{H}_0 \mathcal{H}_3, \nonumber\\
 \mathcal{F}_{21}= &\dfrac{\Sigma^2  \left(r^{2}_{H}+a^2\right)^2}{(r-r_H) \Delta,_r(r_H)  ~\rho^2(r_H)}  (-E+j \Omega+e A_0) \mathcal{H}_0 \mathcal{H}_1 \mathcal{W}', \nonumber\\
 \mathcal{F}_{22}=&-\dfrac{\Sigma^2  \left(r^{2}_{H}+a^2\right)^2}{(r-r_H) \Delta,_r(r_H)  ~\rho^2(r_H)} (-E+j \Omega+e A_0)^2 \mathcal{H}_0^2+ \dfrac{\Delta_\theta}{\rho^2(r_H)} \mathcal{H}_{2}^{2} J_{\theta}^2   \nonumber \\
& +\dfrac{\rho^2(r_H) \Sigma^2}{\Delta_{\theta} \left(r^{2}_{H}+a^2\right)^2 \sin^2\theta} \left(J_\phi +e A_3\right)^2   \mathcal{H}_{3}^2            +m^2, \nonumber\\
 \mathcal{F}_{23}=& -\dfrac{\Delta_\theta}{\rho^2(r_H)} \mathcal{H}_1 \mathcal{H}_2  ~J_{\theta}~ \mathcal{W}', \nonumber\\
  \mathcal{F}_{24}=&-    \dfrac{\rho^2(r_H) ~ \Sigma^2}{\Delta_{\theta} \left(r^{2}_{H}+a^2\right)^2 \sin^2\theta}         \left( J_{\phi}+e A_3\right)~ \mathcal{W}' ~\mathcal{H}_1 \mathcal{H}_3, \nonumber\\
 \mathcal{F}_{31}=&\dfrac{\Sigma^2  \left(r^{2}_{H}+a^2\right)^2}{(r-r_H) \Delta,_r(r_H)  ~\rho^2(r_H)} (-E+j \Omega+e A_0) \mathcal{H}_0 \mathcal{H}_2~ J_{\theta} , \nonumber\\
 \mathcal{F}_{32}=&-\dfrac{(r-r_H) \Delta,_r(r_H)}{\rho^2(r_H)} \mathcal{H}_1 \mathcal{H}_2 ~J_{\theta}~ \mathcal{W}', \nonumber\\
 \mathcal{F}_{33}=& -\dfrac{\Sigma^2  \left(r^{2}_{H}+a^2\right)^2}{(r-r_H) \Delta,_r(r_H)  ~\rho^2(r_H)} (-E+j \Omega+e A_0)^2 ~\mathcal{H}_{0}^2+ \dfrac{(r-r_H) \Delta,_r(r_H)}{\rho^2(r_H)} \mathcal{H}_{1}^2 ~\mathcal{W}'^2 \nonumber\\ & +    \dfrac{\rho^2(r_H) \Sigma^2}{\Delta_{\theta} \left(r^{2}_{H}+a^2\right)^2 \sin^2\theta}   \left(J_\phi +e A_3\right)^2   \mathcal{H}_{3}^2              +m^2,  \nonumber\\
  \mathcal{F}_{34}=&-    \dfrac{\rho^2(r_H) ~ \Sigma^2}{\Delta_{\theta} \left(r^{2}_{H}+a^2\right)^2 \sin^2\theta}         \left( J_{\phi}+e A_3\right)~ J_\theta ~\mathcal{H}_2 \mathcal{H}_3, \nonumber\\
 \mathcal{F}_{41}=&\dfrac{\Sigma^2  \left(r^{2}_{H}+a^2\right)^2}{(r-r_H) \Delta,_r(r_H)  ~\rho^2(r_H)} (-E+j \Omega+e A_0)~ \left( J_{\phi}+e A_3\right) ~ \mathcal{H}_0 \mathcal{H}_3 ,\nonumber\\
 \mathcal{F}_{42}=&-\dfrac{(r-r_H) \Delta,_r(r_H)}{\rho^2(r_H)} \left( J_{\phi}+e A_3\right) \mathcal{H}_1 \mathcal{H}_2 ~ \mathcal{W}',\nonumber\\
 \mathcal{F}_{43}=& -\dfrac{\Delta_\theta}{\rho^2(r_H)}~\left( J_{\phi}+e A_3\right) ~J_{\theta}~ \mathcal{H}_2 \mathcal{H}_3 , \nonumber\\
 \mathcal{F}_{44}=& -\dfrac{\Sigma^2  \left(r^{2}_{H}+a^2\right)^2}{(r-r_H) \Delta,_r(r_H)  ~\rho^2(r_H)} (-E+j \Omega+e A_0)^2 ~\mathcal{H}_{0}^2+ \dfrac{(r-r_H) \Delta,_r(r_H)}{\rho^2(r_H)} \mathcal{H}_{1}^2 ~\mathcal{W}'^2 \nonumber\\ & +   \dfrac{\Delta_\theta}{\rho^2(r_H)} \mathcal{H}_{2}^{2} J_{\theta}^2   +m^2, 
\end{align}
where $\mathcal{W}'=\partial_r \mathcal{W}$, $J_\theta=\partial_\theta \Theta$ and $J_\phi=\partial_\phi \Theta$.

%\begin{align}
%&g^{00}=   - \dfrac{\Sigma^2  \left(r^{2}_{H}+a^2\right)^2}{(r-r_H) \Delta,_r(r_H)  ~\rho^2(r_H)}
%\nonumber\\ 
%& g^{11}=    \dfrac{(r-r_H) \Delta,_r(r_H)}{\rho^2(r_H)} 
%\nonumber\\ 
% g^{22}=      \dfrac{\Delta_\theta}{\rho^2(r_H)}
%\nonumber\\ 
%& g^{33}=      \dfrac{\rho^2(r_H) \Sigma^2}{\Delta_{\theta} \left(r^{2}_{H}+a^2\right)^2 \sin^2\theta}
%\end{align}

Eq. \eqref{eqn 50} has a non-trivial solution if the determinant of the matrix $\mathcal{F}$ equals zero. If $det(\mathcal{F})=0$, then eq. \eqref{eqn 50} gives

\begin{align} \label{b}
&\mathcal{O}(\beta^6) (\partial_r\mathcal{W})^{18} +\mathcal{O}(\beta^5) (\partial_r\mathcal{W})^{16}+ \mathcal{O}(\beta^4) (\partial_r\mathcal{W})^{14}+ \mathcal{O}(\beta^3) (\partial_r\mathcal{W})^{12}  +\mathcal{O}(\beta^2) \nonumber\\
& (\partial_r\mathcal{W})^{10} +\mathcal{O}(\beta^2) (\partial_r\mathcal{W})^{8}+ \mathcal{O}(\beta^2) (\partial_r\mathcal{W})^{6}+ \mathcal{O}(\beta^2) (\partial_r\mathcal{W})^{4}  +\mathcal{O}(\beta^2) (\partial_r\mathcal{W})^{2} + \mathcal{O}(\beta^2) \nonumber\\
&+ \left[  A^{*}_0  (\partial_r\mathcal{W})^2  +A^{*}_1\right] \left\lbrace B_0 +B_2 (\partial_r\mathcal{W})^{2}+ B_4 (\partial_r\mathcal{W})^{4} +B_6  (\partial_r\mathcal{W})^{6} \right\rbrace ~=~0.
\end{align}

(The expressions for $A_i$ and $B_i$ are given in Appendix \ref{appendix A}.)

Solving eq. \eqref{b} by neglecting the higher order terms of $\beta$, we obtain the solution of the derivative of radial action as
\begin{align}\label{eqn 53}
\partial_r \mathcal{W}= & \Biggl( -\dfrac{m^2   \rho^2(r_H) +J_{\theta}^2 \Delta_\theta}{(r-r_H) \Delta,_r(r_H)} +\dfrac{(-E+j \Omega+e A_0)^2 \left(r^{2}_{H}+a^2\right)^2 \Sigma^2  }{(r-r_H)^2 \Delta^{2},_r(r_H)}	\nonumber\\
& -\dfrac{\left(J_\phi +e A_3\right)^2 \rho^4(r_H) 	~			\Sigma^2 ~\csc^2\theta }{(r-r_H) \left(r^{2}_{H}+a^2\right) \Delta_\theta ~\Delta,_r(r_H) }	\Biggr)^{\dfrac{1}{2}}	~\times \left(1+\dfrac{\chi_1}{\chi_2} \beta \right).
\end{align}
(The expressions of $\chi_1$ and $\chi_2$ are given in Appendix \ref{appendix B}.)
%   (r-r_H)^3 			\left(r^{2}_{H}+a^2\right)^6 	
						
%	\Delta^{3}_\theta~ 					\Delta^{3},_r(r_H)~ \rho^6(r_H)
%
%		\left(J_\phi +e A_3\right)^3 			\rho^2(r_H) 				\Sigma^4 ~\csc^4\theta

%		(-E+j \Omega+e A_0)^2 
The imaginary part of the radial action is obtained by integrating eq. \eqref{eqn 53} at the pole, $r=r_H$ as

\begin{align}
Im \mathcal{W}_\pm= & \pm Im \bigintss dr ~\Biggl(  -\dfrac{m^2   \rho^2(r_H) +J_{\theta}^2 \Delta_\theta}{(r-r_H) \Delta,_r(r_H)} +\dfrac{(-E+j \Omega+e A_0)^2 \left(r^{2}_{H}+a^2\right)^2 \Sigma^2  }{(r-r_H)^2 \Delta^{2},_r(r_H)}	\nonumber\\
		&~~~~~~ ~~~~~~~~~~~~~-\dfrac{\left(J_\phi +e A_3\right)^2 \rho^4(r_H) 	~			\Sigma^2 ~\csc^2\theta }{(r-r_H) \left(r^{2}_{H}+a^2\right) \Delta_\theta ~\Delta,_r(r_H) }	\Biggr)^{\dfrac{1}{2}}	~\times \left(1+\dfrac{\chi_1}{\chi_2} \beta \right) \nonumber\\
 =&	\pm \dfrac{\pi  (E -j \Omega-e A_0) \left(r^{2}_{H}+a^2\right) \left(1+\frac{\Lambda a^2}{3} \right)}{2 \left[ (r_H-M)-\frac{\Lambda~ r_H}{3} \left(2 r_H^{2}+a^2 \right) \right]} ~ \left( 1+ \beta ~ \Xi	\right)	,	 
\end{align}
where
\begin{align}
\Xi=&	-\dfrac{4 \left(J_\phi +e A_3\right)^2 \rho^2(r_H)~ 				\Sigma^2 ~\csc^2\theta}{\left(r^{2}_{H}+a^2\right)^2 \Delta_\theta} + 		 \dfrac{1}{2~ \rho^4(r_H) 	~			\Sigma^2 \left(J_\phi +e A_3\right)\left(J_\phi +e A_3-1\right)}		\nonumber\\
& \Biggl[ 3m^2 ~\left(r^{2}_{H}+a^2\right)^2 ~\Delta_\theta ~\sin^{2}\theta \bigl\{5 J^{2}_\theta \Delta_\theta+ 3 m^2~\rho^2(r_H) 	\bigr\} +\left(J_\phi +e A_3\right)^3 			\rho^2(r_H) 	~			\Sigma^2 \nonumber\\
& \biggl[ -8 J_{\theta}^2~ \Delta_\theta \left(J_\phi +e A_3-1\right) +m^2 ~\rho^2(r_H) \bigl\{ 8+7\left(J_\phi +e A_3\right) \bigr\} \biggr] \Biggr].
\end{align}

%   (r-r_H)^3 			\left(r^{2}_{H}+a^2\right)^6 	
						
%	\Delta^{3}_\theta~ 					\Delta^{3},_r(r_H)~ \rho^6(r_H)
%
%		\left(J_\phi +e A_3\right)^3 			\rho^2(r_H) 				\Sigma^4 ~\csc^4\theta

%		(-E+j \Omega+e A_0)^2 

Here $\mathcal{W}_+$ and $\mathcal{W}_-$ indicate the radial action of the outgoing and ingoing particles respectively. According to WKB approximation, the tunneling probabilities are given by
\begin{align}
& \mathcal{P}_{outgoing}= \exp \biggl[ -2 \biggl\{Im (\mathcal{R}_+) + Im(V) \biggr\} 		\biggr] \nonumber\\
& \text{and} \nonumber\\
& \mathcal{P}_{ingoing}=\exp \biggl[ -2 \biggl\{Im (\mathcal{R}_-) + Im(V) \biggr\} 		\biggr].
\end{align}

There is a $100\%$ probability of ingoing particle to enter the black hole in accordance with the semiclassical WKB approximation. Thus, the tunneling rate of $W^+$ boson particles is given by
\begin{align}
\Gamma_{rate}=\dfrac{\mathcal{P}_{outgoing}}{\mathcal{P}_{ingoing}}= &\exp\bigl[		-4 	\{Im (\mathcal{R}_+)\bigr] \nonumber\\
=& \exp \Biggl[		 \dfrac{-2~\pi  (E -j \Omega-e A_0) \left(r^{2}_{H}+a^2\right) \left(1+\frac{\Lambda a^2}{3} \right)}{ \left[ (r_H-M)-\frac{\Lambda~ r_H}{3} \left(2 r_H^{2}+a^2 \right) \right]} ~ \left( 1+ \beta ~ \Xi	\right)	\Biggr].
\end{align}
The Boltzman factor gives the Hawking temperature of the black hole \cite{kerner2}. Thus, the GUP modified Hawking temperature is derived as
\begin{align}
T_{d4}=& \dfrac{ \left[ (r_H-M)-\frac{\Lambda~ r_H}{3} \left(2 r_H^{2}+a^2 \right) \right]} {2~\pi   \left(r^{2}_{H}+a^2\right) \left(1+\frac{\Lambda a^2}{3} \right)}~ \left( 1- \beta ~ \Xi	\right)	\nonumber\\
=& ~T_o~ \left( 1- \beta ~ \Xi	\right),
\end{align}
where $T_o=\dfrac{ \left[ (r_H-M)-\frac{\Lambda~ r_H}{3} \left(2 r_H^{2}+a^2 \right) \right]}{2~\pi   \left(r^{2}_{H}+a^2\right) \left(1+\frac{\Lambda a^2}{3} \right)} $, is the original Hawking temperature of KNdS black hole without any quantum gravity effects. The modified Hawking temperature $T_{d4}$ may be lower or greater than the original Hawking temperature $T_{o}$ according to $\Xi>0$ or $\Xi<0$ respectively. The modified Hawking temperature depends on the quantum numbers (mass and angular momentum) of the emitted vector boson particles.

The modified heat capacity is calculated as 
\begin{align} \label{eqn 59}
C_{H4}=&\left(\dfrac{\partial M}{\partial r_h} \right) \left( \dfrac{\partial r_h}{\partial T_d} \right) \nonumber\\
=& \dfrac{2 \pi \left(a^2+r^2\right)^2 \left(3+a^2 \Lambda\right) \left[ a^2 \left( 3+r^2 \Lambda \right)+3 \left(Q^2-r^2+r^4 \Lambda \right)\right]}{3 \left[ a^4 \left(r^2 \Lambda -3\right) +3 r^2 \left( r^2+r^4 \Lambda-3 Q^2\right)+a^2 \left( 8 r^4 \Lambda -3 Q^2-12 r^2 \right) 			\right]} \nonumber \\
& \times \left( 1+ \beta ~\Xi \right).
\end{align}
From eq. \eqref{eqn 59}, it is observed that the modified heat capacity reduces to the original heat capacity when $\beta=0$. Thus, the original heat capacity is recovered in the absence of the quantum gravity effects. The modified heat capacity $C_{H4}$ is higher or lower than the original heat capacity $C_{o}$, according to $\Xi>0$ or $\Xi<0$ respectively.

\section{Remnant of 3-dimensional KNdS black hole}

Many studies have shown that the GUP effect could give a black hole remnant \cite{chen1, chen2, myung1, gangopadhyay1, feng1,medved,bargueno}. \textbf{For 3-dimensional  KNdS black hole the quantum gravity effects slow down the increase of  Hawking temperature. This leads to the formation of remnants in black hole evaporation.} From this point, we will investigate the remnant of 3-dimensional  KNdS black hole. The tunneling particle's mass is no longer considered in the following discussion since the tunneling particles at the event horizon are effectively massless. According to the uncertainty principle, the lower limit of the tunneling particle energy can be expressed \cite{adler,amelino} as
\begin{align}
E\geq \dfrac{\hbar}{\Delta x}.
\end{align}
Near the event horizon, one may take the uncertainty of the position as the radius of the black hole \cite{adler,amelino} as
\begin{align}
\Delta x  \approx r_{BH}=r_H.
\end{align}
From eq. \eqref{eq 34}, we obtain
\begin{align}\label{eq 37}
T_d=& \frac{ \left[ r_H-\dfrac{1}{2} \left(1-\dfrac{\Lambda r^{2}_{H}}{3} \right) \left(r_H+\dfrac{a^2}{r_H} \right) -\dfrac{Q^2}{2 r_H}-\frac{\Lambda~ r_H}{3} \left(2 r_H^{2}+a^2 \right) \right]}{2\pi   \left(r^{2}_{H}+a^2\right) \left(1+\frac{\Lambda a^2}{3} \right)} \nonumber\\
& \times \left(	1- \frac{4  \beta J_{\theta}^{2} \Delta_{\theta}}{r^{2}_H+ a^2 \cos^{2}\theta}		\right).
\end{align}
From eq. \eqref{eq 37}, it is observed that when
\begin{align}
r_H \leq \sqrt{\dfrac{4 \beta J^{2}_{\theta} \left(1+ \Lambda a^2 \cos^{2}\theta \right)-3 a^2 \cos^{2}\theta}{3}},
\end{align}
the modified Hawking temperature becomes negative. This violates the law of black hole thermodynamics and thus has no physical meaning. It is clear that the evaporation will stop under the effects of GUP. Thus the Hawking temperature becomes zero when $r_H$ reaches the minimum radius, $r_{min}$ as
\begin{align}
r_{min}=& \sqrt{\dfrac{4 \beta J^{2}_{\theta} \left(1+ \Lambda a^2 \cos^{2}\theta \right)-3 a^2 \cos^{2}\theta}{3}}.
\end{align}
% \nonumber\\
%=& l_p \sqrt{\dfrac{1}{3} \left[ \dfrac{4 \beta_o J^{2}_\theta}{\hbar^2} \left(1+ \Lambda a^2 \cos^{2}\theta \right)- \dfrac{3 a^2 \cos^{2}\theta}{l^{2}_p} \right]}
Using eq. \eqref{eq r} in eq. \eqref{eq 34}, we obtain the expression of $T_d$ in terms of the mass of the black hole as

\begin{align}
T_d \approx \dfrac{\zeta_1}{\zeta_2}  \left(	1- \frac{4  \beta J_{\theta}^{2} \Delta_{\theta}}{\dfrac{1}{\alpha^{2}_1} \left( 1+\dfrac{4 \Lambda M^2}{3 \beta_{1}^2 \alpha_1} 	\right)^2 \left( M+ \sqrt{M^2-(a^2+Q^2)\alpha_1}	\right)^2+ a^2 \cos^{2}\theta}		\right),
\end{align}
where
\begin{align}
\zeta_1= & \dfrac{1}{\alpha_1} \left( 1+\dfrac{4 \Lambda M^2}{3 \beta_{1}^2 \alpha_1} 	\right) \left( M+ \sqrt{M^2-(a^2+Q^2)\alpha_1}	\right) 
\nonumber\\
&\left[ 1 -\dfrac{\Lambda a^2}{3}	-\dfrac{2 \Lambda}{3 ~\alpha^{2}_1} \left( 1+\dfrac{4 \Lambda M^2}{3 \beta_{1}^2 \alpha_1} 	\right)^2 \left( M+ \sqrt{M^2-(a^2+Q^2)\alpha_1}	\right)^2 \right]-M,  \nonumber\\
\zeta_2= & 2 \pi \left(1+ \dfrac{\Lambda a^2}{3} \right)  \left[ a^2 + \dfrac{1}{\alpha^{2}_1} \left( 1+\dfrac{4 \Lambda M^2}{3 \beta_{1}^2 \alpha_1} \right)^2 \left( M+ \sqrt{M^2-(a^2+Q^2)\alpha_1}	\right)^2		\right].
\end{align}

To make the Hawking temperature $T\geq 0$ i.e. to ensure the GUP corrected temperature has a physical meaning, the mass of the black hole must hold the inequality
\begin{align}
M \geq & \dfrac{\beta^{2}_{1}}{8 (a^2+Q^2) \Lambda-6 \beta^{2}_{1}} 	\left[ 3 \sqrt{-\alpha_1 (a^2+Q^2)}  + \dfrac{3}{2} \biggl\{-4 \alpha_1 (a^2+Q^2)  \right.  \nonumber\\ &  \left. + \dfrac{8 \alpha_1  \zeta_3 \left\lbrace3 \beta^{2}_1-4 (a^2+Q^2) \Lambda \right\rbrace	\left( -a^2+Q^2+4 J^{2}_\theta \alpha_1 \beta \Delta_\theta - a^2 \alpha_1 \cos^{2}\theta \right)		}{3 \beta^{2}_1}		\biggr\}^{\frac{1}{2}}	\right].
\end{align} 
% \left( -a^2+Q^2+4 J^{2}_\theta \alpha_1 \beta \Delta_\theta - a^2 \alpha_1 \cos^{2}\theta \right)
%where $\zeta_3=-a^2+Q^2+4 J^{2}_\theta \alpha_1 \beta \Delta_\theta - a^2 \alpha_1 \cos^{2}\theta$

It is noted that the mass of the black hole has a minimum value which is given by
\begin{align}
M_{min} = & \dfrac{\beta^{2}_{1}}{8 (a^2+Q^2) \Lambda-6 \beta^{2}_{1}} 	\left[ 3 \sqrt{-\alpha_1 (a^2+Q^2)}  + \dfrac{3}{2} \biggl\{-4 \alpha_1 (a^2+Q^2)  \right.  \nonumber\\ &  \left. + \dfrac{8 \alpha_1  \zeta_3 \left\lbrace3 \beta^{2}_1-4 (a^2+Q^2) \Lambda \right\rbrace	\left( -a^2+Q^2+4 J^{2}_\theta \alpha_1 \beta \Delta_\theta - a^2 \alpha_1 \cos^{2}\theta \right)		}{3 \beta^{2}_1}		\biggr\}^{\frac{1}{2}}	\right]. 
\end{align}

\section{Graphical Analysis}
In this section, we will examine graphically the effects of parameters $\beta$, $\Lambda$ and $m$ on the modified Hawking temperatures and modified heat capacities with respect to event horizon $r_H$.
\subsection{Temperature $T_{d}$ with radius of event horizon $r_{H}$ for 3-dimensional KNdS black hole}
This subsection is devoted to analysing the behaviour of modified Hawking temperature. The parameters are taken as follows $a=0.3$, $Q=1$, $\theta=\frac{\pi}{2}$ and $J_\theta=0.1$.

%\begin{figure}[h!]
%\footnotesize
%\stackunder[5pt]{\includegraphics[width=160pt]{Temperature3drb2.pdf}}{}%
%\hspace{1cm}%
%\stackunder[5pt]{\includegraphics[width=160pt]{Temperature3drb1.pdf}}{}
%\caption{Hawking temperature $T_d$ with respect to radius of event horizon $r_H$ for different values of  $\beta$.}
%\label{fig 1}
%\end{figure}
\begin{figure}[h!]
\centering % \begin{center}/\end{center} takes some additional vertical space
\includegraphics[width=210pt]{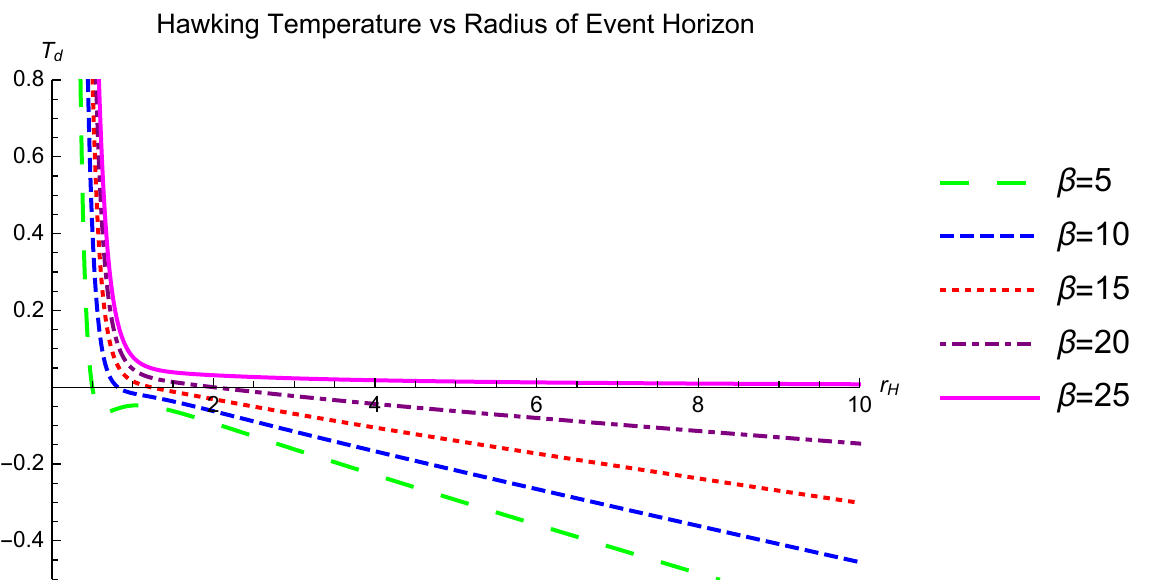}
\hfill
\includegraphics[width=210pt]{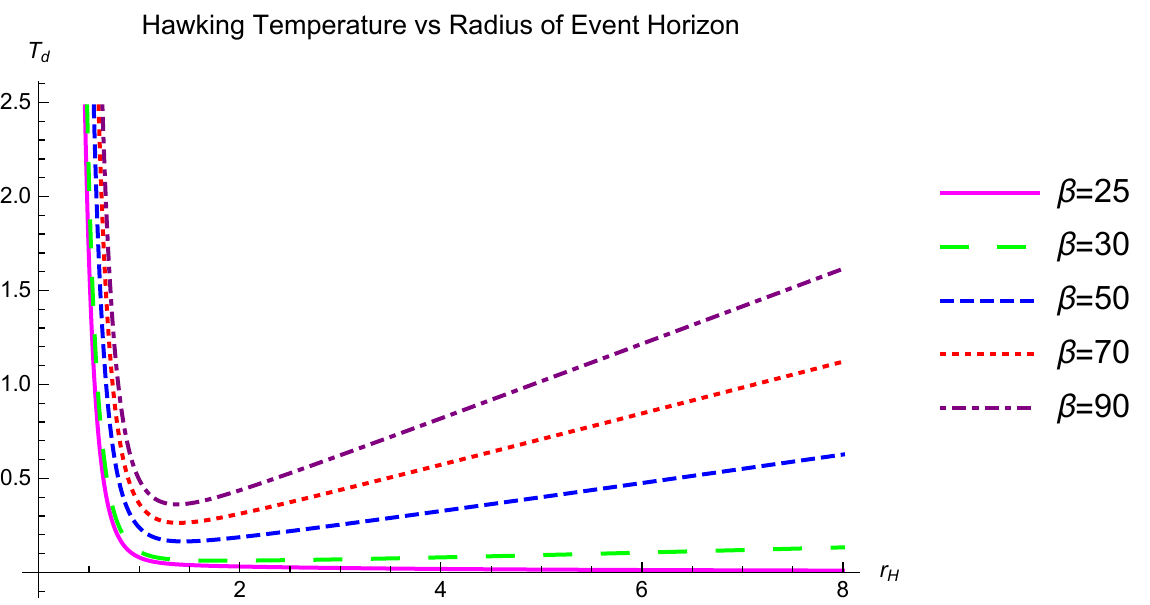}
% "\includegraphics" is very powerful; the graphicx package is already loaded
\caption{\label{fig 1} Hawking temperature $T_d$ with respect to radius of event horizon $r_H$ for different values of  $\beta$.}
\end{figure}

%\caption{\label{fig:i} Always give a caption.}

%\begin{figure}[h!]
%    \centering
%    \subfigure[]{\includegraphics[width=0.45\textwidth]{Temperature3drL.pdf}} \hfill
%    \subfigure[]{\includegraphics[width=0.45\textwidth]{Temperature3drm.pdf}} 
%    \caption{(a) blah (b) blah (c) blah (d) blah}
%\end{figure}
\begin{figure}[h!]
\centering
  \centerline{\includegraphics[width=300pt]{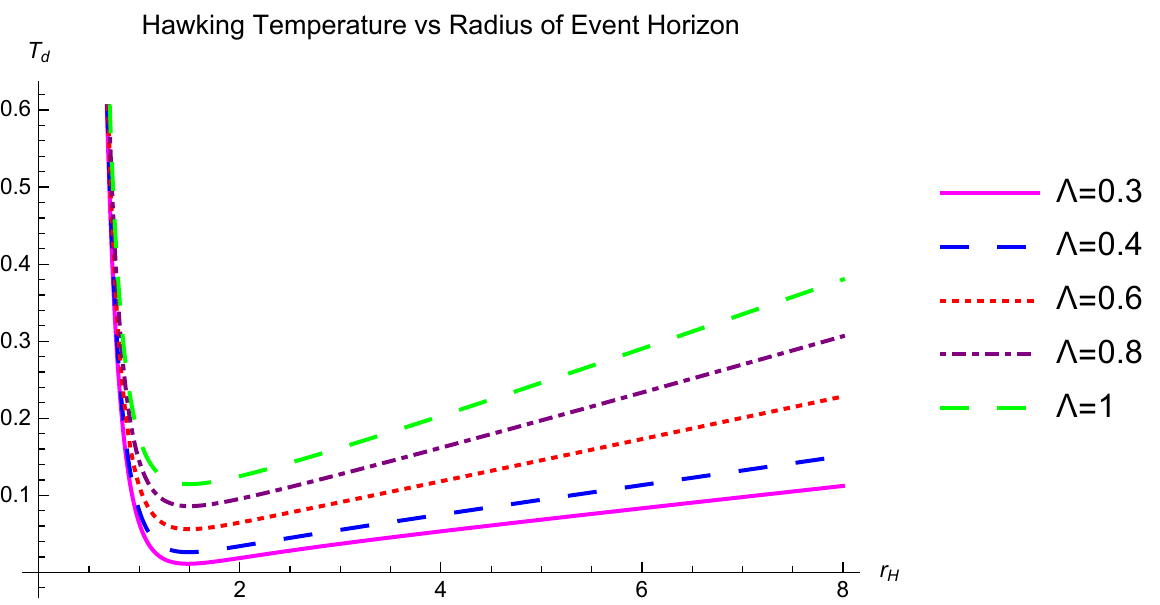}}
  \caption{ \label{fig 2} Hawking temperature $T_d$ with respect to the radius of event horizon $r_H$ for different values of cosmological constant $\Lambda$.} 
\end{figure}
\begin{enumerate}
\item  Figure \ref{fig 1} indicates the variation of the modified Hawking temperature with $r_H>0$, for different values of $\beta$  with fixed value of the cosmological constant $\Lambda=1$ and $m=0.1$. We observe that for $\beta=25$, the temperature decreases and tends to zero. As the radius of horizon increases for $\beta<25$, the modified Hawking temperatures become negative. This negative temperature and divergent behaviour reveals the nonphysically unstable state of the black hole \cite{wajiha1}. Moreover, for $\beta>25$, as the horizon increases, the temperature decreases and once the minimum value is reached, the temperature increases.
%after attaining the minimum value, the temperature increases. 
It is worth mentioning that, for $\beta<25$, we observe the nonphysical behaviour with negative temperature and for $\beta=25$, the temperature vanishes. Furthermore, the temperatures are positive when $\beta>25$. The
$\beta$ effects decelerate the increase in Hawking temperature which is also shown numerically in table \ref{table 1}.

\item Figure \ref{fig 2} shows the behaviour of modified Hawking temperature with $r_H>0$, for different values of positive cosmological constant $\Lambda>0$. The parameters are taken as follows: $m=0.1$ and $\beta=40$. It is observed that the modified Hawking temperatures decrease and attain its minimum values, which is also calculated numerically in table \ref{table 2}. Later, it keeps on increasing as $r_H$ increases and its behaviour is linear. The change in $\Lambda$ gives the diverging temperature $T_d$ as $r_H$ increases.
\begin{figure}[h!]
\centering
  \centerline{\includegraphics[width=300pt]{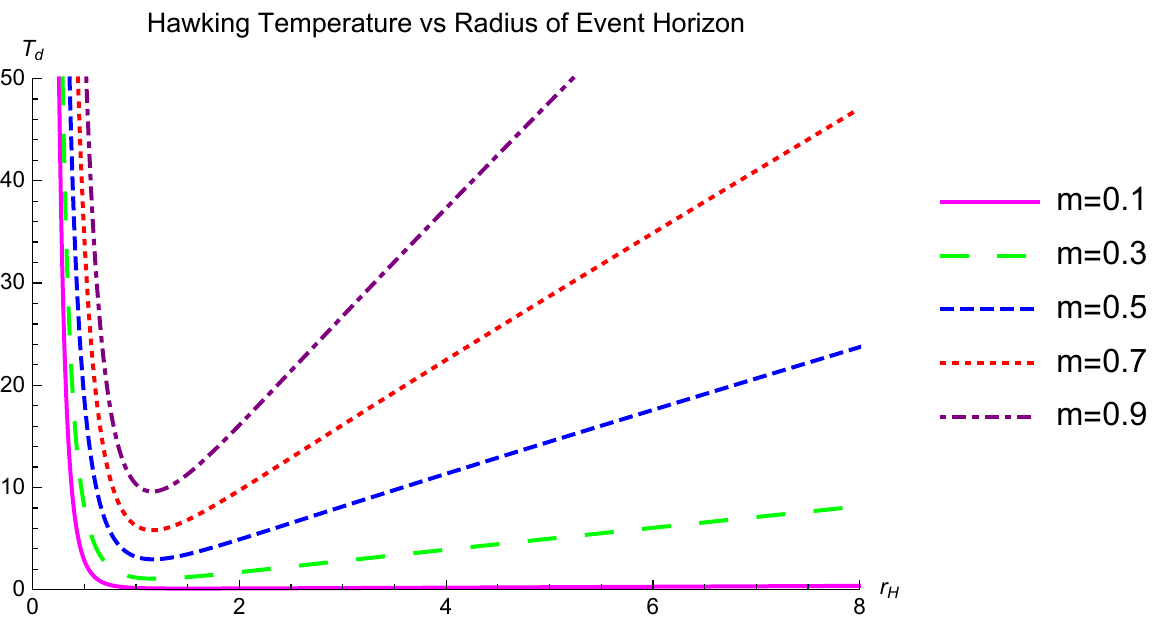}}
  \caption{Hawking temperature $T_d$ with respect to radius of event horizon $r_H$ for different values of $m$.} 
  \label{fig 3}
\end{figure}
\item Figure \ref{fig 3} indicates the behaviour of Hawking temperature for different values of $m$. At first, the Hawking temperature drops suddenly and after attaining its minimum point, it keeps on increasing with increasing the horizon radius. Increasing the values of parameter $m$ tend to increase the modified Hawking temperatures of the black hole which is also shown numerically in table \ref{table 3}.
\begin{figure}[h!]
\centering
  \centerline{\includegraphics[width=300pt]{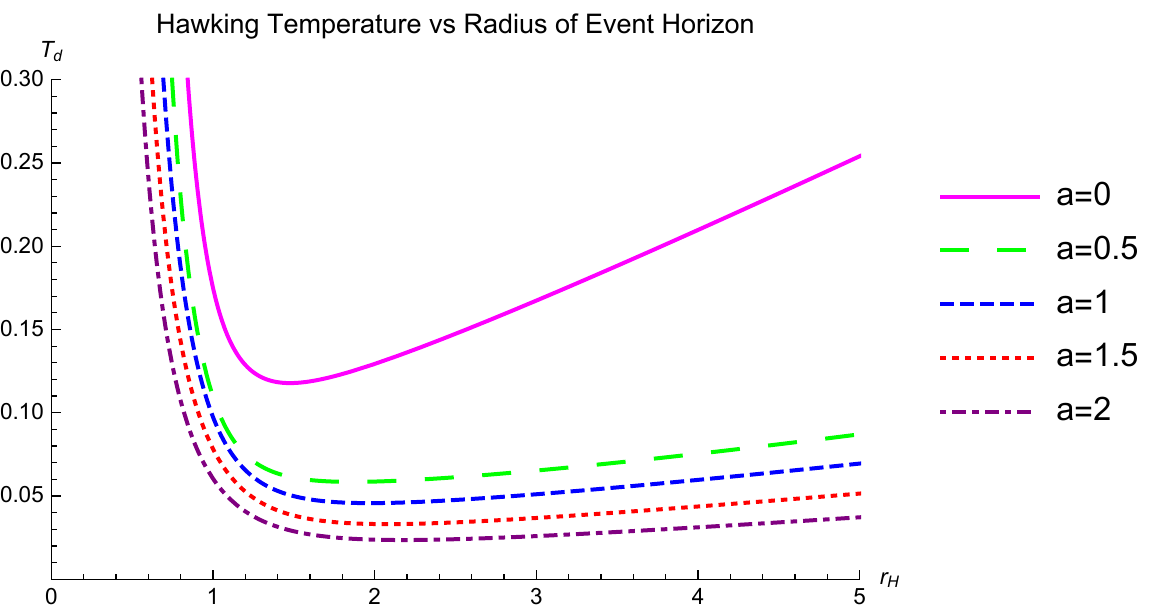}}
  \caption{Hawking temperature $T_d$ versus radius of event horizon $r_H$ for different values of $a$.} 
  \label{fig 4}
\end{figure}
\item Figure \ref{fig 4} provides the graphical analysis of $T_d $ via horizon radius $r_H$ for different values of spin parameter $a$. The pink line, which corresponds to $a=0$ represents the Hawking temperature graph for Reissner-Nordstrom-de Sitter (RNdS) black hole. The graph shows that the Hawking temperature of the RNdS black hole is greater than that of the KNdS black hole. The effect of spin parameter $a$ decelerates the increase in Hawking temperature. This graphical presentation is compatible with the numerical calculation of table \ref{table 4}.
\end{enumerate}

From figures \ref{fig 1} to \ref{fig 4} show that the temperature cools down to the minimum $T_{d}^{C}$, at $r_{H}=r_{H}^{C}$ and keeps on increasing as $r_{H}>r_{H}^{C}$.

\begin{table} [tbp]
%\tbl{The tabulated values of the critical radius $r_{H}^{C}$ and the critical temperature $T_{H}^{C}$ for different values of $\beta$.}
\centering
\begin{tabular}{|M{1.7 cm}|M{1.7 cm}|M{1.7 cm}|M{1.7 cm}|M{1.7 cm}|M{1.7 cm}|}
\hline
\multicolumn{6}{c}{a=0.3; \quad \quad m=0.1;\quad \quad $\Lambda=1$}  \\
\hline
$\beta$ & 25 & 30 & 50 & 70 & 90 \\
\hline
$r_{H}^{c}$ & 7.97272 & 1.81347 & 1.43917 & 1.3935 & 1.37473 \\
\hline
$T_{d}^{c}$ & 0.00953 & 0.06208 & 0.16458 & 0.26337 &  0.36169 \\
\hline
\end{tabular}
\caption{\label{table 1} The tabulated values of the critical radius $r_{H}^{C}$ and the critical temperature $T_{H}^{C}$ for different values of $\beta$.}
\end{table}

\begin{table} [!htbp]
%\tbl{The tabulated values of the critical radius $r_{H}^{C}$ and the critical temperature $T_{H}^{C}$ for different values of $\Lambda$. }
{\begin{tabular}{|M{1.7 cm}|M{1.7 cm}|M{1.7 cm}|M{1.7 cm}|M{1.7 cm}|M{1.7 cm}|}

\hline
\multicolumn{6}{c}{a=0.3; \quad \quad m=0.1;\quad \quad $\beta=40$}  \\
\hline
$\Lambda$ & 0.3 & 0.4 & 0.6 & 0.8 & 1 \\
\hline
$r_{H}^{c}$ & 1.4842 & 1.4876  & 1.49087 & 1.49664 & 1.5016\\
\hline
$T_{d}^{c}$ & 0.01115 & 0.02619 & 0.05600 & 0.08596 & 0.11457 \\
\hline
\end{tabular}}
\caption{\label{table 2} The tabulated values of the critical radius $r_{H}^{C}$ and the critical temperature $T_{H}^{C}$ for different values of $\Lambda$.}
\end{table}
\FloatBarrier
\begin{table} [!htbp]
%\tbl{The tabulated values of the critical radius $r_{H}^{C}$ and the critical temperature $T_{H}^{C}$ for different values of emitted particle mass $m$. }
{\begin{tabular}{|M{1.7 cm}|M{1.7 cm}|M{1.7 cm}|M{1.7 cm}|M{1.7 cm}|M{1.7 cm}|}

\hline
\multicolumn{6}{c}{a=0.3; \quad \quad $\Lambda=1$;\quad \quad $\beta=40$}  \\
\hline
$m$ & 0.1 & 0.3 & 0.5 & 0.7 & 0.9 \\
\hline
$r_{H}^{c}$ & 1.5016 & 1.18474 & 1.16561 & 1.16027 & 1.15807\\
\hline
$T_{d}^{c}$ & 0.11457 & 1.07833 & 2.97447 & 5.81695 & 9.60655 \\
\hline
\end{tabular}}
\caption{\label{table 3}	The tabulated values of the critical radius $r_{H}^{C}$ and the critical temperature $T_{H}^{C}$ for different values of emitted particle mass $m$.	}
\end{table}

\begin{table} [h !]
%\tbl{The tabulated values of the critical radius $r_{H}^{C}$ and the critical temperature $T_{H}^{C}$ for different values of $a$. }
{\begin{tabular}{|M{1.7 cm}|M{1.7 cm}|M{1.7 cm}|M{1.7 cm}|M{1.7 cm}|M{1.7 cm}|}

\hline
\multicolumn{6}{c}{m=0.1; \quad \quad $\Lambda=1$;\quad \quad $\beta=40$}  \\
\hline
$a$ & 0.3 & 0.4 & 0.6 & 0.8 & 1 \\
\hline
$r_{H}^{c}$ & 1.44254 & 1.86004 & 1.96988 & 2.07671 & 2.17653\\
\hline
$T_{d}^{c}$ & 0.11791 & 0.05845 & 0.04575 & 0.03314 & 0.02359 \\
\hline
\end{tabular}}
\caption{\label{table 4}	The tabulated values of the critical radius $r_{H}^{C}$ and the critical temperature $T_{H}^{C}$ for different values of $a$.	}
\end{table}

\subsection{Heat capacity $C_H$ with radius of event horizon $r_{H}$ for 3-dimensional KNdS black hole}
This subsection focuses on analysing modified heat capacity in different domains of event horizon radius $r_H$ with fixed parameters:  $a=0.3$, $Q=1$, $\theta=\frac{\pi}{2}$ and $J=0.1$. The heat capacity is connected to the local thermal stability. If the black hole has a negative heat capacity, it is unstable to thermal radiation and if it has a positive heat capacity, it is stable to thermal radiation. 

\begin{enumerate}[(i)]

\begin{figure}[h!]
\centering
  \centerline{\includegraphics[width=300pt]{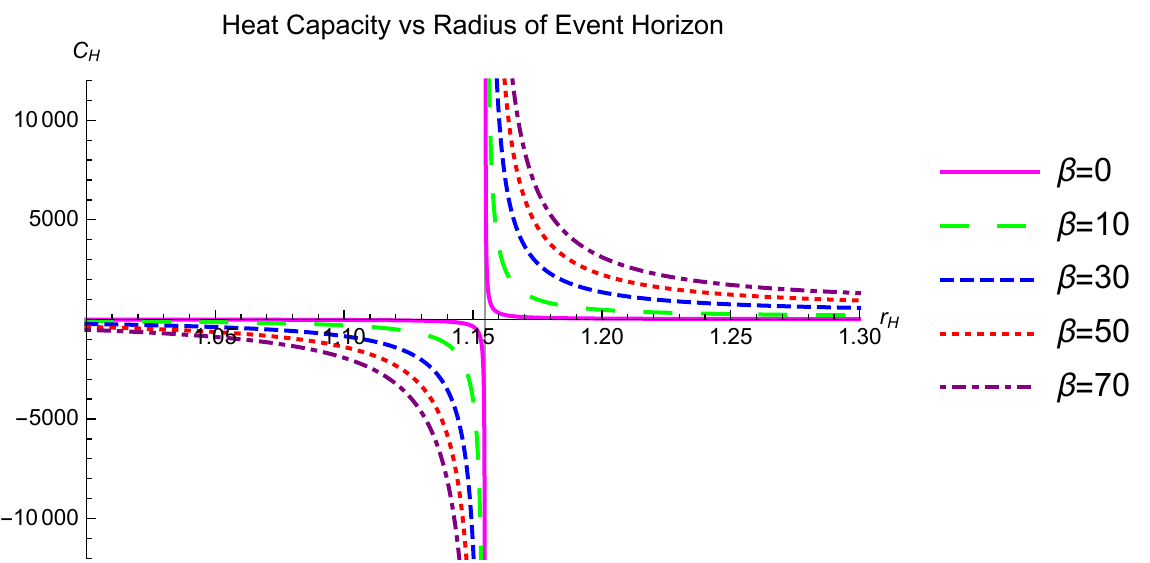}}
  \caption{Heat capacity $C_H$ versus radius of event horizon $r_H$ for different values of $\beta$.} 
  \label{fig 5}
\end{figure}
\item For fixed parameters $a=0.3$, $\Lambda=1$ and $m=0.5$, the variation of heat capacity $C_H$ with the horizon radius $r_{H}$ for different values of $\beta$ is shown in figure \ref{fig 5}. It is observed that the phase transition occurs at $r_{H}=r_{H}^*=1.15467$, $r_{H}^*$ denotes the position of the horizon radius at which the phase transition takes place. The position of phase transition remains unchanged in 3-dimensional KNdS black hole for different values of $\beta$. 
\FloatBarrier
\begin{figure}[h!]
\centering
  \centerline{\includegraphics[width=300pt]{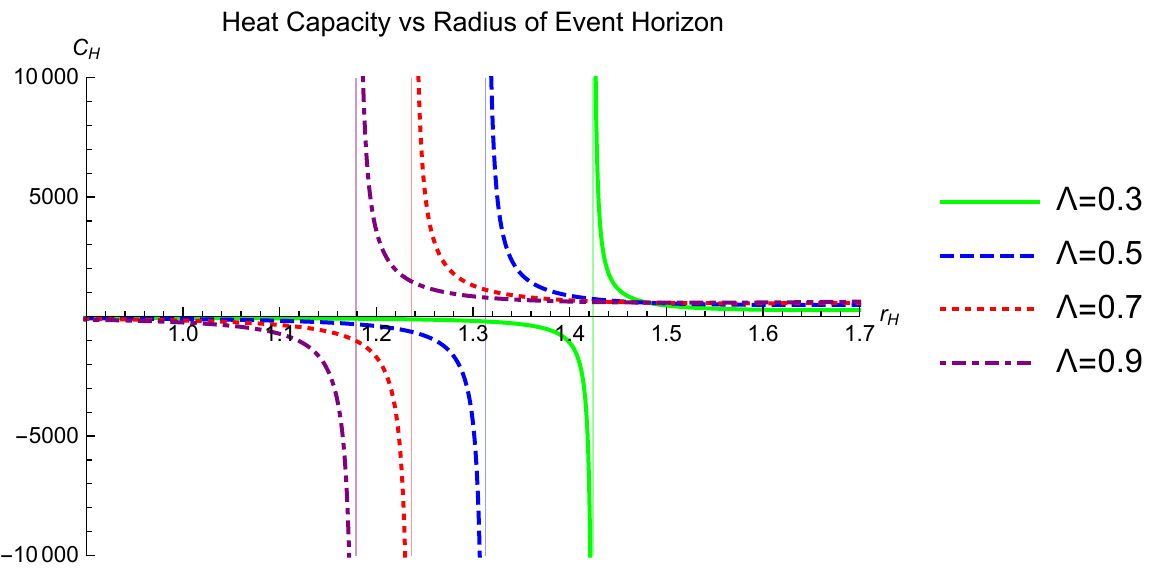}}
  \caption{Heat capacity $C_H$ versus radius of event horizon $r_H$ for different values of $\Lambda$.} 
  \label{fig 6}
\end{figure}
\FloatBarrier
\item Figure \ref{fig 6} illustrates the behaviour of $C_H$ w.r.t $r_{H}$  for fixed parameters $a=0.3$, $\beta=40$, $m=0.5$ and for different values of $\Lambda$. For different values of cosmological constant $\Lambda$, there are different positions of phase transition for 3-dimensional KNdS black hole. Increasing the values of $\Lambda$, the positions of phase transition $r_{H}^*$ are shifted towards the origin, which ensures the black hole faster stability for larger value of $\Lambda$.  Different positions of phase transition for different black hole parameters are presented in table \ref{table 5}.
\begin{figure}[h!]
\centering
  \centerline{\includegraphics[width=300pt]{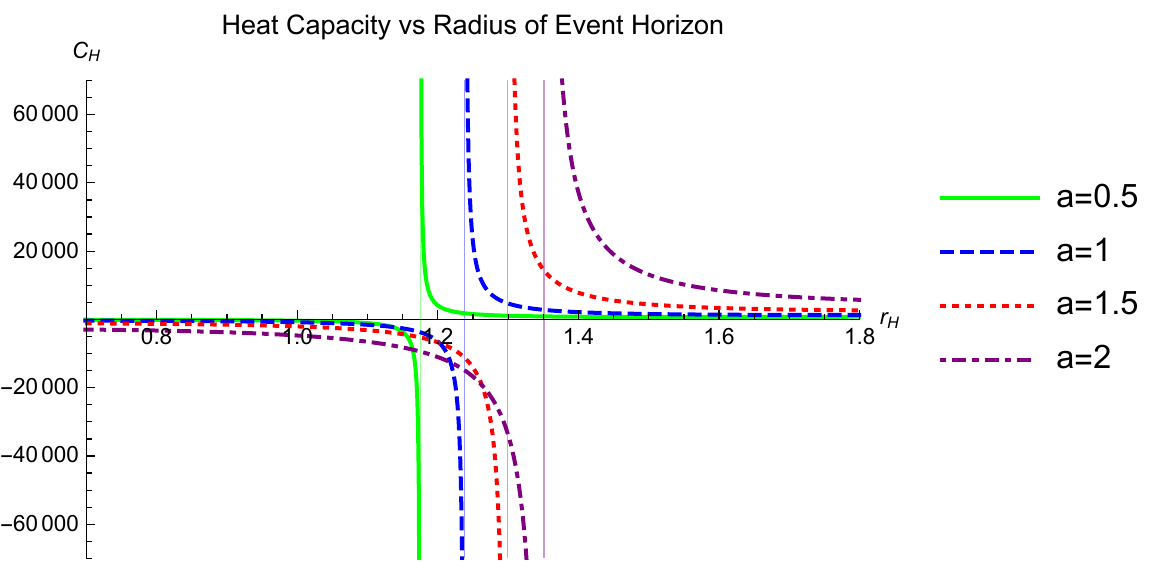}}
  \caption{Heat capacity $C_H$ versus radius of event horizon $r_H$ for different values of $a$.} 
  \label{fig 7}
\end{figure}
\item Figure \ref{fig 7} shows the variation of $C_H$ for different values of spin parameter $a$ and for fixed $\Lambda=1$, $\beta=40$, $m=0.5$. Increasing the value of rotation parameter $a$, the phase transition occurs at larger value of horizon radius $r_{H}^*$ which is also shown in table \ref{table 6}. Moreover, larger the value of $a$ delays the stability of the black hole.
 \par Figures \ref{fig 5} to \ref{fig 7} show that there is a phase transition when $r_{H}=r_{H}^{*}$. The positions of phase transition are shown in tables \ref{table 5} and \ref{table 6}. The black holes are unstable in the region $0< r_{H}< r_{H}^{*}$ and stable in the region $r_{H}^{*} < r_{H}<  \infty$ i.e. the smaller black holes are less stable than larger black holes and vice versa.
\end{enumerate}
\FloatBarrier
\begin{table}  [!htbp]
%\tbl{The tabulated values of the position of phase transition $r_{H}^*$ for different values of $\Lambda$. }
{\begin{tabular}{|M{1.7 cm}|M{1.7 cm}|M{1.7 cm}|M{1.7 cm}|M{1.7 cm}|}

\hline
\multicolumn{5}{c}{a=0.3; \quad \quad $m=0.5$;\quad \quad $\beta=40$}  \\
\hline
$\Lambda$ & 0.3 & 0.5 & 0.7 & 0.9  \\
\hline
$r_{H}^{*}$ & 1.42403 &  1.31282 & 1.23652   &  1.17887  \\
\hline
\end{tabular}}
\caption{\label{table 5}	The tabulated values of the position of phase transition $r_{H}^*$ for different values of $\Lambda$.	}
\end{table}
\FloatBarrier

\begin{table} [h !]
%\tbl{The tabulated values of the position of phase transition $r_{H}^*$ for different values of $a$. }
{\begin{tabular}{|M{1.7 cm}|M{1.7 cm}|M{1.7 cm}|M{1.7 cm}|M{1.7 cm}|}

\hline
\multicolumn{5}{c}{$\Lambda=0.1$; \quad \quad $m=0.5$;\quad \quad $\beta=40$}  \\
\hline
$a$ & 0.5 & 1 & 1.5 & 2  \\
\hline
$r_{H}^{*}$ & 1.17485 &  1.23815 & 1.29861   &  1.35098  \\
\hline
\end{tabular}}
\caption{\label{table 6} The tabulated values of the position of phase transition $r_{H}^*$ for different values of $a$.	}
\end{table}

\subsection{ Hawking temperature $T_{d4}$ with radius of event horizon for 4-dimensional KNdS black hole}
This subsection is devoted to analysing the behaviour of modified Hawking temperatures of KNdS black hole. The parameters are taken as follows: $Q=0.1$, $\theta=\frac{\pi}{2}$, $e=1$, $J_\theta=0.1$ and $J_\phi=0.2$.

\begin{enumerate}[(i)]
\FloatBarrier
\begin{figure}[h!]
\centering
  \centerline{\includegraphics[width=300pt]{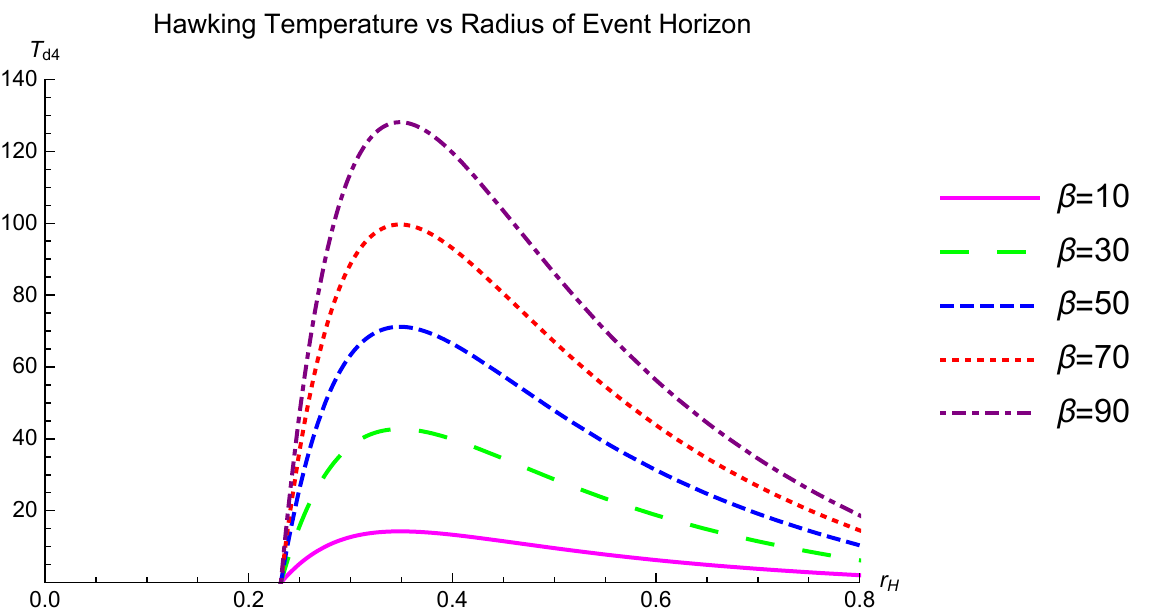}}
  \caption{Hawking temperature $T_{d4}$ with respect to radius of event horizon $r_H$ for different values of $m$.} 
  \label{fig 8}
\end{figure}

\item  Figure \ref{fig 8} shows the behaviour of $T_{d4}$ for different values of $\beta$ and for fixed values of $a=0.2$, $\Lambda=1$ and $m=0.1$. $T_{d4}$ increases exponentially and reaches its maximum height $T_{d4}^{max}$ for different values of $\beta$. Further, the temperature keeps on decreasing with increasing the horizon radius. It is noteworthy to mention that the temperature increases with increasing the values of $\beta$. Hence, the $\beta$ effects accelerate the increase in $T_{d4}$. The proof of the above statement is also calculated in table \ref{table 7}. 
\FloatBarrier
\begin{figure}[h!]
\centering
  \centerline{\includegraphics[width=300pt]{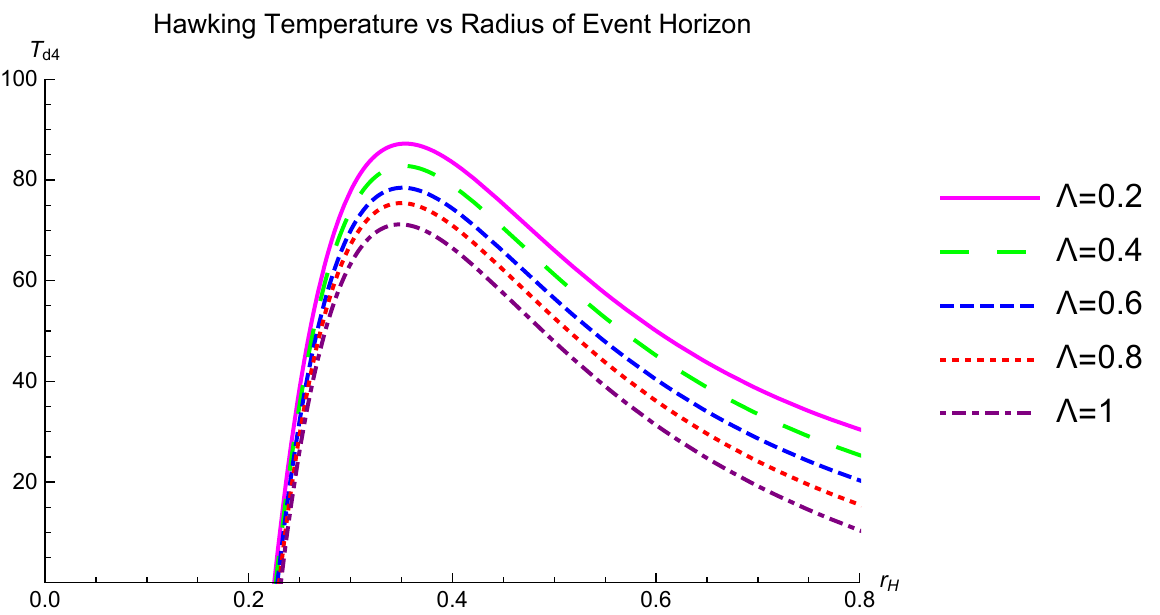}}
  \caption{Hawking temperature $T_{d4}$ with respect to radius of event horizon $r_H$ for different values of $\beta$.} 
  \label{fig 9}
\end{figure}
\item The behaviour of $T_{d4}$ for varying cosmological constant $\Lambda$ is depicted in figure \ref{fig 9}. The parameters are as follows: $a=0.2$, $\beta=50$ and $m=0.1$. The temperature increases to a certain height and attains its peak point $T_{d4}^{~max}$ at $r_{H}=r_{H}^c$, then $T_{d4}$ decreases as $r_{H}$ increases. It is noted that the $\Lambda$ effects decelerate the increase in $T_{d4}$ which is also shown numerically in table \ref{table 8}.

\item The behaviour of $T_{d4}$ w.r.t $r_H$ for varying mass of the vector boson particle $m$ is depicted in Figure \ref{fig 10}. The temperature increases to a certain height $T_{d4}^{~max}$ and then decreases with increasing $r_H$. The rate of increase of temperature $T_{d4}$ is dependent on the increase of $m$. The validity of the above statements is calculated numerically in table \ref{table 9}.
\item Figure \ref{fig 11} shows the behaviour of $T_{d4}$ w.r.t $r_{H}$ for varying $a$ and fixed  $\beta=50$, $\Lambda=1$ and $m=0.1$. The temperatures increase exponentially upto $T_{d4}^{~max}$ and decrease with increasing the horizon radius. The temperature gradually increases with decreasing the values of rotation parameter, $a$. The numerical calculation shown in table \ref{table 10} supports the above statement.
\FloatBarrier
\begin{figure}[h!]
\centering
  \centerline{\includegraphics[width=300pt]{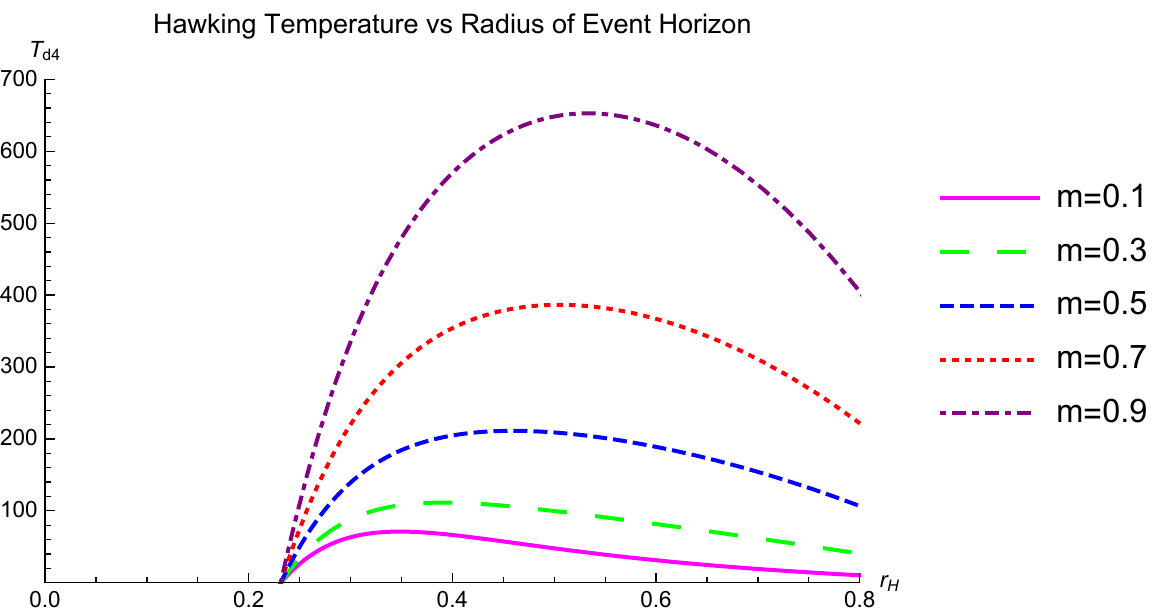}}
  \caption{Hawking temperature $T_{d4}$ with respect to radius of event horizon $r_H$ for different values of $m$.} 
  \label{fig 10}
\end{figure}
\FloatBarrier
\begin{figure}[h!]
\centering
  \centerline{\includegraphics[width=300pt]{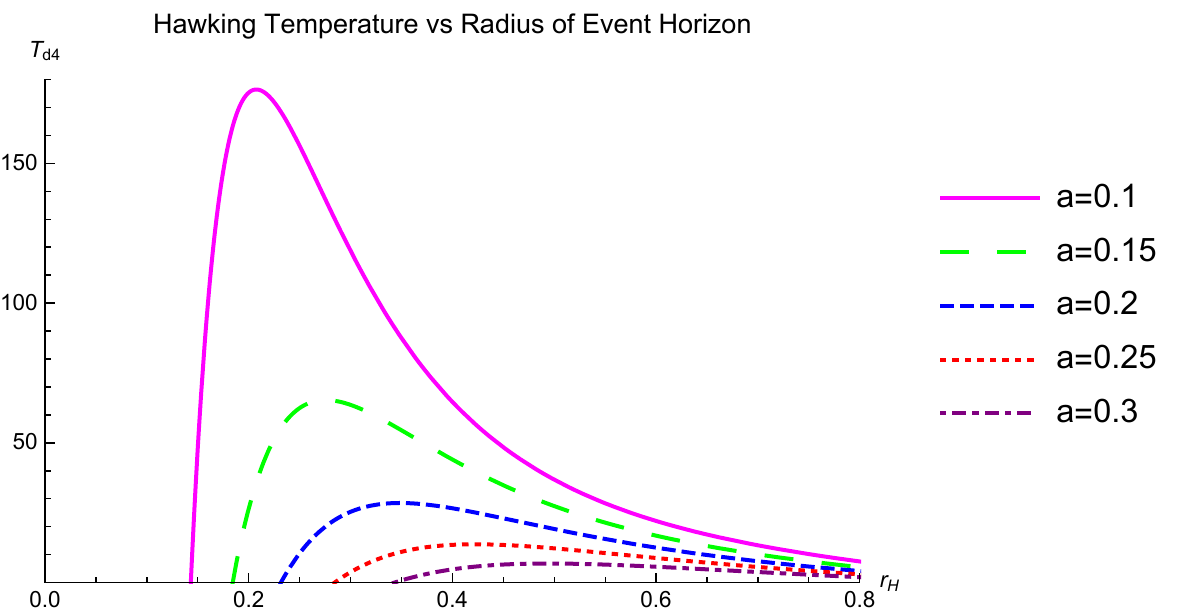}}
  \caption{Hawking temperature $T_{d4}$ with respect to radius of event horizon $r_H$ for different values of $a$.} 
  \label{fig 11}
\end{figure}
\FloatBarrier
\end{enumerate}
We present the following tables to see the effects of $\beta$, $\Lambda$, $m$ and $a$ on the Hawking temperature. Table \ref{table 7} confirms that $T_{d4}^{~max}$ gradually increases on increasing $\beta$ (for fixed values of $\lambda$, $m$ and $a$). Similarly table \ref{table 9} also confirms that $T_{d4}^{~max}$ gradually increases on increasing $m$ (for fixed values of $\beta$, $\lambda$, and $a$).
% Tables \ref{table 7} and \ref{table 9} confirm that $T_{d4}^{~max}$ gradually increase on increasing $r_{H}$. 
On the contrary, tables \ref{table 8} and \ref{table 10} confirm that $T_{d4}^{~max}$ gradually decreases with increasing $\lambda$ (for fixed values of $\beta$, $m$ and $a$) and $a$ (for fixed values of  $\beta$, $\lambda$, and $m$) respectively. From table \ref{table 9}, it is observed that $T_{d4}^{~max}$ is highly dependent on $m$ compared to that of $\beta$, $\Lambda$ and $a$.

\FloatBarrier
\begin{table} [!htbp]
%\tbl{The tabulated values of $r_{H}^{C}$ and $T_{H}^{~max}$ for different values of  $\beta$. }
{\begin{tabular}{|M{1.7 cm}|M{1.7 cm}|M{1.7 cm}|M{1.7 cm}|M{1.7 cm}|M{1.7 cm}|}
\hline
\multicolumn{6}{c}{$a$=0.2; \quad \quad $\Lambda=1$;\quad \quad $m=0.1$}  \\
\hline
$\beta$ & 10 & 30 & 50 & 70 & 90 \\
\hline
$r_{H}^{c}$ & 0.34861	& 0.34849	& 0.34847	& 0.34846	& 0.34845\\
\hline
$T_{d4}^{~max}$ & 14.2974	& 42.7385	& 71.1796	& 99.6206	& 128.062 \\
\hline
\end{tabular}}
\caption{\label{table 7}	The tabulated values of $r_{H}^{C}$ and $T_{d4}^{~max}$ for different values of  $\beta$.	}
\end{table}

\FloatBarrier
\begin{table} [!htbp]
%\tbl{The tabulated values of $r_{H}^{C}$ and $T_{H}^{~max}$ for different values of  $\Lambda$. }
{\begin{tabular}{|M{1.7 cm}|M{1.7 cm}|M{1.7 cm}|M{1.7 cm}|M{1.7 cm}|M{1.7 cm}|}
\hline
\multicolumn{6}{c}{$a$=0.2; \quad \quad $\Lambda=1$;\quad \quad $\beta=50$}  \\
\hline
$\Lambda$ & 0.2 & 0.4 & 0.6 & 0.8 & 1 \\
\hline
$r_{H}^{c}$ &  0.35376	& 0.35244	& 0.35117	& 0.34979	& 0.34847	\\
\hline
$T_{d4}^{~max}$ & 87.2216	& 82.793		& 78.4615	& 75.4107	& 71.1796	 \\
\hline
\end{tabular}}
\caption{\label{table 8}	The tabulated values of $r_{H}^{C}$ and $T_{d4}^{~max}$ for different values of  $\Lambda$.	}
\end{table}

\FloatBarrier
\begin{table} [!htbp]
%\tbl{The tabulated values of $r_{H}^{C}$ and $T_{H}^{~max}$ for different values of  $m$. }
{\begin{tabular}{|M{1.7 cm}|M{1.7 cm}|M{1.7 cm}|M{1.7 cm}|M{1.7 cm}|M{1.7 cm}|}
\hline
\multicolumn{6}{c}{$a=0.2$; \quad \quad $\Lambda=1$;\quad \quad $\beta=50$}  \\
\hline
$m$ & 0.1 & 0.3 & 0.5 & 0.7 & 0.9 \\
\hline
$r_{H}^{c}$ & 0.34847	& 0.39069	& 0.45936	& 0.50492	& 0.53291 \\
\hline
$T_{d4}^{~max}$ & 71.1796	& 111.502	& 211.269	& 386.449	& 652.623 \\
\hline
\end{tabular}}
\caption{\label{table 9}	The tabulated values of $r_{H}^{C}$ and $T_{d4}^{~max}$ for different values of  $m$.	}
\end{table}

\FloatBarrier
\begin{table} [!htbp]
%\tbl{The tabulated values of $r_{H}^{C}$ and $T_{H}^{~max}$ for different values of  $a$. }
{\begin{tabular}{|M{1.7 cm}|M{1.7 cm}|M{1.7 cm}|M{1.7 cm}|M{1.7 cm}|M{1.7 cm}|}
\hline
\multicolumn{6}{c}{$\beta=20$; \quad \quad $\Lambda=1$;\quad \quad $m=0.1$}  \\
\hline
$a$ & 0.1 & 0.15 & 0.2 & 0.25 & 0.3 \\
\hline
$r_{H}^{c}$ & 0.2075		& 0.27555	& 0.34852	& 0.42236	& 0.49421 \\
\hline
$T_{d4}^{~max}$ & 176.342	& 65.3181	& 28.518		& 13.7598	& 6.89087 \\
\hline
\end{tabular}}
\caption{\label{table 10}	The tabulated values of $r_{H}^{C}$ and $T_{d4}^{~max}$ for different values of  $a$.	}
\end{table}

\subsection{Heat Capacity $C_{H4}$ versus horizon radius $r_H$ for 4-dimensional KNdS black hole}
This subsection studies the modified heat capacity for different values of  $\beta$, $\Lambda$ and $a$ for fixed values of $Q=1$, $\theta=\frac{\pi}{2}$, $e=1$, $J_\theta=0.1$ and $J_\phi=0.2$.

\begin{enumerate}[(i)]
\item The variation of modified heat capacity for different values of $\beta$ is shown in figure \ref{fig 12}. The parameters are taken as $a=0.2$, $\Lambda=0.5$ and $m=0.1$. It is observed that there is one position of phase transition in the absence of GUP, but there are two positions of phase transition under the influence of GUP. The first phase transition in figure \ref{fig 12} is due to the quantum gravity effects and it occurs at $r_{H}=r_{H_1}=0.99338$. The position of second phase transition is at $r_{H}=r_{H_2}=1.29667$. The variation of $\beta$ doesn't affect the position of phase transition.
%It is worth mentioning that KNdS black hole has only one phase transition  quantum gravity effects but it has two phase transitions under the influence of GUP. The quantum gravity effect do not affect these positions of the phase transition. 
 
\item For $a=0.2$, $\beta=15$ and $m=0.1$, the variation of $C_{H4}$ w.r.t $r_H$ changing the values of $\Lambda$ is illustrated in figure \ref{fig 13}. Varying the values of cosmological constant $\Lambda$, the positions of phase transition are also varied. With increasing the values of $\Lambda$, the position of phase transition is shifted toward the origin, which implies a slower rate of becoming a stable black hole. Table \ref{table 11} is constructed numerically to show the different positions of phase transitions. 
\begin{figure}[h!]
\centering
  \centerline{\includegraphics[width=300pt]{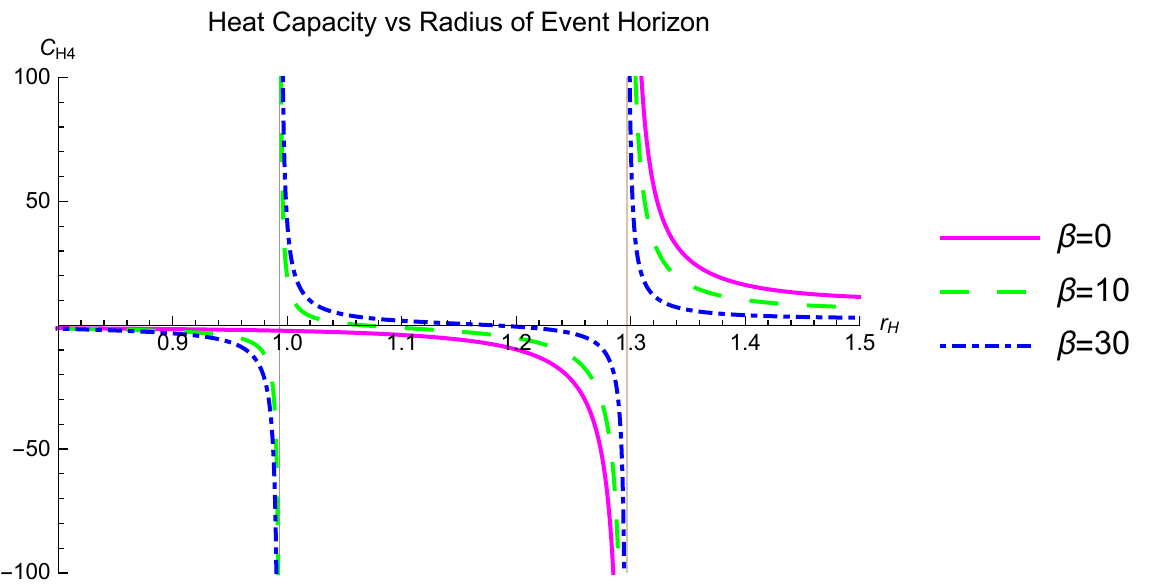}}
  \caption{Heat Capacity $C_{H4}$ with respect to radius of event horizon $r_H$ for different values of $\beta$.} 
  \label{fig 12}
\end{figure}
\FloatBarrier

\begin{figure}[h!]
\centering
  \centerline{\includegraphics[width=300pt]{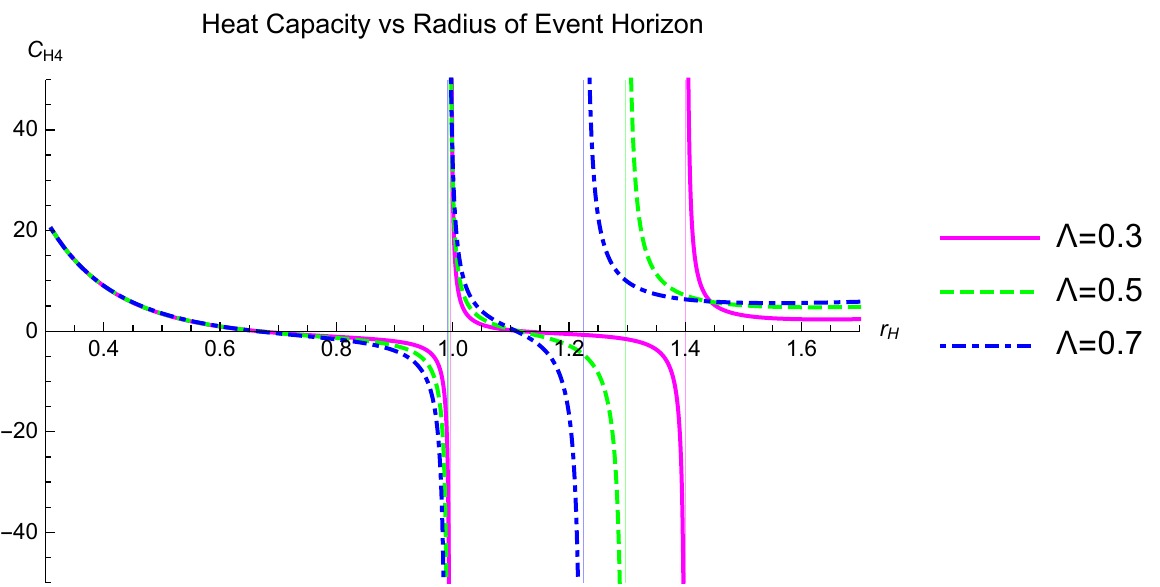}}
  \caption{Heat Capacity $C_{H4}$ with respect to radius of event horizon $r_H$ for different values of $\Lambda$.} 
  \label{fig 13}
\end{figure}

\item Figure \ref{fig 14} represents the behaviour of $C_{H4}$ versus $r_{H}$ for fixed $\beta=15$, $\Lambda=0.5$ and $m=0.1$. The position of phase transition is shifted to far away from the origin toward the positive direction of $r_H$ with increasing the value of spin parameter $a$. It shows that the black hole becomes stable faster with increasing the value of rotational parameter $a$ which is also indicated numerically by table \ref{table 11}.
\FloatBarrier
\begin{figure}[h!]
\centering
  \centerline{\includegraphics[width=300pt]{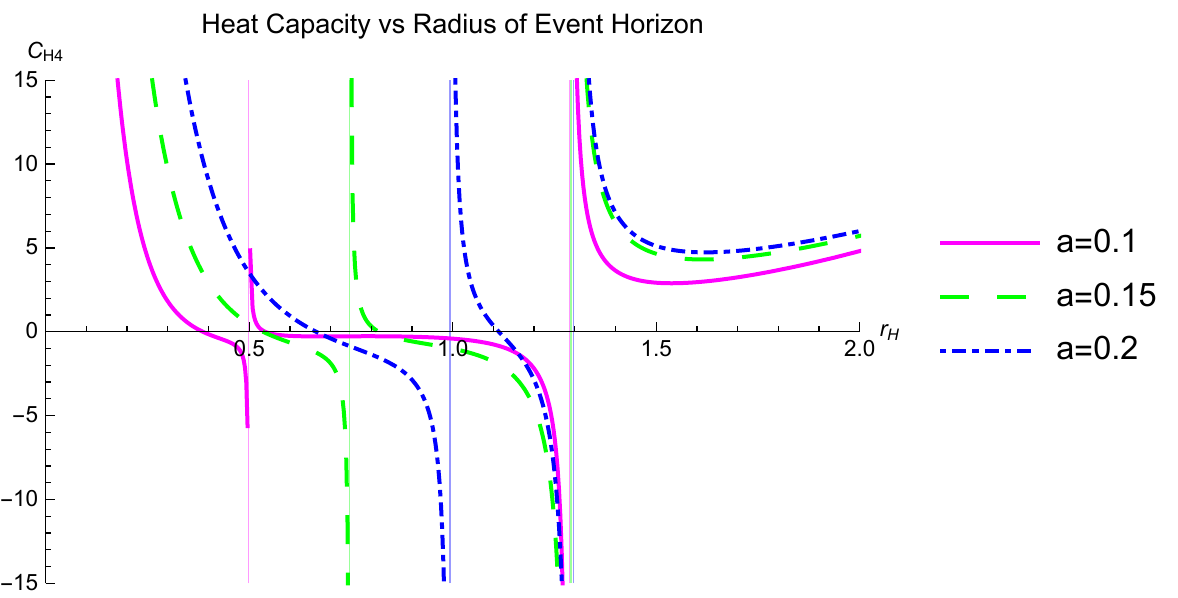}}
  \caption{Heat Capacity $C_{H4}$ with respect to radius of event horizon $r_H$ for different values of $a$.} 
  \label{fig 14}
\end{figure}
\FloatBarrier
\end{enumerate}
\FloatBarrier
\begin{table} [!htbp]
%\tbl{The tabulated values of $r_{H_1}$ and $r_{H_2}$ for different values of  $\Lambda$ and $a$. }
{\begin{tabular}{|M{1.5 cm}|M{1.5 cm}|M{1.5 cm}||M{1.5 cm}|M{1.5 cm}|M{1.5 cm}|}
\hline
\multicolumn{3}{|c||}{$a=0.2$; \quad \quad $\beta=15$;\quad \quad $m=0.1$}&  \multicolumn{3}{c|}{$\Lambda=0.5$; \quad \quad $\beta=15$;\quad \quad $m=0.1$}  \\
\hline
$\Lambda$ &  $r_{H_1}$	& $r_{H_2}$ & $a$ & $r_{H_1}$ & $r_{H_2}$ \\
\hline
0.3 	& 	0.99602	& 1.40094	& 0.1	 & 0.49917 & 1.2864\\
\hline
0.5 	& 0.99338	& 1.29667	& 0.4	& 0.7472  & 1.29073 \\
\hline
0.7 	& 0.99075	& 1.22479	& 0.6	& 0.99338 & 1.29667 \\
\hline
\end{tabular}}
\caption{\label{table 11} The tabulated values of $r_{H_1}$ and $r_{H_2}$ for different values of  $\Lambda$ and $a$.}
\end{table}

\section{Conclusion}

This work studies the GUP effects on tunneling of massive vector boson particles from KNdS black hole. Firstly, we utilize the GUP-corrected Lagrangian of massive vector field and derive the modified wave equation for massive vector boson particles. Using the modified wave equation, the quantum tunneling of KNdS black hole in 3-dimensional and 4-dimensional frame dragging coordinates are investigated. Further, the modified Hawking temperatures and heat capacities due to GUP are derived. They depend not only on the quantum gravity parameter $\beta$, spin parameter $a$, mass of the emitted particle $m$, cosmological constant $\Lambda$, charge of the black hole $Q$ but also on angular coordinates $\theta$, $J_\theta$ and $J_\phi$. \textbf{For 3-dimensional KNdS black hole, the corrected Hawking temperature is lower than the original Hawking temperature which shows that the quantum gravity effects slow down the increase of the Hawking radiation temperature. Moreover, the modified heat capacity of 3-dimensional KNdS black hole is greater than the original heat capacity. In the case of 4-dimensional KNdS black hole, the modified Hawking temperature is either lower or higher than the original Hawking temperature according to $\Xi>0$ or $\Xi<0$ respectively. Further, the modified heat capacity is either lower or higher than the original Hawking temperature according to $\Xi<0$ or $\Xi>0$ respectively.} The stable and unstable formation of black hole are studied in quantum gravity effects. The remnant of 3-dimensional KNdS black hole is also discussed in the presence of quantum gravity effects. We also illustrate the graphs of modified Hawking temperatures and heat capacities and explore the effects of $\beta$, $\Lambda$, $a$ and $m$. \textbf{ If the radius of event horizon $r_H$ increases, the modified Hawking temperature of a 3-dimensional KNdS black hole tends to decrease for $\beta<25$, but for $\beta>25$, the temperature cools down till it reaches its minimum point and then increases, which leads to the formation of stable black hole}. For a 4-dimensional KNdS black hole with the above fixed parameters, the modified Hawking temperature increases w.r.t $r_H$ and after attaining maximum height, the temperature eventually goes down. It is worth noting that there are one phase transition and two phase transitions for a non-zero horizon of 3-dimensional KNdS black hole and 4-dimensional KNdS black hole respectively. Different positions of phase transitions are due to the quantum gravity effects. It is noted that for different values of dimensionless parameter $\beta$, the position of phase transitions remain the same in 3-dimensional and 4-dimensional KNdS black hole under the influence of quantum gravity effects. The modified Hawking temperatures and heat capacities tend to the original Hawking temperature and heat capacity of KNdS black hole in the absence of quantum gravity effects. Hence quantum gravity effects modified the Hawking temperature and heat capacity of the black hole.

\appendix
\section{Coefficients of eq. \eqref{b}}
\label{appendix A}
%\section*{Appendix A (Coefficients of the differential equation \eqref{b})}

\small{
\begin{align}
A^{*}_0=& \dfrac{(r-r_{H} )\Delta,_r(r_H)}{ \rho^2(r_H) }, \\
A^{*}_1=& m^2+\dfrac{J_{\theta}^2 \Delta_\theta}{\rho^2(r_H)}- \dfrac{(-E+j \Omega+e A_0)^2  \left(r^{2}_{H}+a^2\right)^2 \Sigma^2}{(r-r_H) \Delta,_r(r_H) \rho^2(r_H) }+  \dfrac{\left(J_\phi +e A_3\right)^2  \rho^2(r_H) \Sigma^2 \csc^2\theta  }{\left(r^{2}_{H}+a^2\right)^2 \Delta_\theta}, \\
B_6= & \dfrac{6 m^2 \beta~ (r-r_{H})^3 ~ \Delta^2,_r(r_H)}{\rho^6(r_H)}, \\
B_4= & \dfrac{ m^2 ~ (r-r_{H})^2 ~ \Delta^2,_r(r_H)}{\rho^4(r_H)} + \dfrac{2 \beta ~ (r-r_{H})^2 ~ \Delta^2,_r(r_H) }{\rho^4(r_H)}   \Biggr[  
 	-\Biggl\{	\dfrac{1}{(r-r_{H}) ~ \Delta,_r(r_H)~ \rho^2(r_H)}	   \nonumber\\ &   \Bigg(	(-E+j \Omega+e A_0)^2   \left(r^{2}_{H}+a^2\right)^2 \Sigma^2        \bigg[-\dfrac{2 \left(J_\phi +e A_3-1\right)\left(J_\phi +e A_3\right)		\rho^2(r_H) \Sigma^2 \csc^2\theta	}{\left(r^{2}_{H}+a^2\right)^2 \Delta_\theta} \nonumber\\
 	& +3 m^2
 	 \biggr] 																					\Bigg)	\Biggr\}		 +m^2   \biggl[	3m^2+ \dfrac{3 J^2_{\theta} \Delta_\theta}{\rho^2(r_H)}	+\dfrac{\left(J_\phi +e A_3\right)\left(2+J_\phi +e A_3\right)		\rho^2(r_H) \Sigma^2 \csc^2\theta}{\left(r^{2}_{H}+a^2\right)^2 \Delta_\theta}	\biggr]	\Biggr], \\
 	B_2= & 	\dfrac{(r-r_{H}) ~ \Delta,_r(r_H)}{\rho^2(r_H)} \Biggl[	\dfrac{(-E+j \Omega+e A_0)^2  \left(J_\phi +e A_3\right)\left(J_\phi +e A_3-1\right) \Sigma^4 \csc^2\theta}{(r-r_{H}) ~ \Delta^2,_r(r_H)}	 \nonumber\\ &  	+2m^4+ \dfrac{2 ~J_{\theta}^2 m^2 \Delta_\theta}{\rho^2(r_H)}   +\dfrac{\left(J_\phi +e A_3\right)\left(1+J_\phi +e A_3\right)m^2 		\rho^2(r_H) \Sigma^2 \csc^2\theta}{\left(r^{2}_{H}+a^2\right)^2 \Delta_\theta}	\nonumber\\ &   	-\dfrac{2 m^2(-E+j \Omega+e A_0)^2   \left(r^{2}_{H}+a^2\right)^2 \Sigma^2    }{(r-r_{H}) ~ \Delta,_r(r_H)~ \rho^2(r_H)} 					+ 2 \beta \Biggl\{ 	\dfrac{1}{(r-r_{H})^2 ~ \Delta^{2},_r(r_H)~ \rho^4(r_H)}	\nonumber\\
 	&	\biggl[	(-E+j \Omega+e A_0)^4   \left(r^{2}_{H}+a^2\right)^4 \Sigma^4	\biggl\{   \dfrac{\left(J_\phi +e A_3-1\right)\left(J_\phi +e A_3\right)		\rho^2(r_H) \Sigma^2 \csc^2\theta}{\left(r^{2}_{H}+a^2\right)^2 \Delta_\theta}	 \nonumber\\		& -3m^2	\biggr\} \biggr]		+  	\dfrac{(-E+j \Omega+e A_0)^2  \left(J_\phi +e A_3\right)^3 \left(J_\phi +e A_3-1\right) \rho^2(r_H) \Sigma^6 \csc^4\theta}{(r-r_{H}) ~\left(r^{2}_{H}+a^2\right)^2 \Delta^{2}_\theta~ \Delta,_r(r_H)}	
 	\nonumber\\	&+ m^2  \bigg[  \dfrac{3 J^{4}_\theta  \Delta_{\theta}^2}{\rho^4(r_H)}	+	\dfrac{\left(J_\phi +e A_3\right)^3 \left\lbrace 1+2 \left(J_\phi +e A_3\right)\right\rbrace \rho^4(r_H) \Sigma^4 \csc^4\theta}{\left(r^{2}_{H}+a^2\right)^4 \Delta^{2}_\theta} \biggr]	\Biggr\}		\Biggr], \\
 B_0=& 	- \Biggl[ \dfrac{J_{\theta}^2 \Delta_\theta}{\rho^2(r_H)}- \dfrac{(-E+j \Omega+e A_0)^2  \left(r^{2}_{H}+a^2\right)^2 \Sigma^2}{(r-r_H) \Delta,_r(r_H) \rho^2(r_H) }+  \dfrac{\left(J_\phi +e A_3\right)^2  \rho^2(r_H) \Sigma^2 \csc^2\theta  }{\left(r^{2}_{H}+a^2\right)^2 \Delta_\theta}		\nonumber\\	
 &	+m^2 		\Biggl] 
 +
 \dfrac{2~\beta}{(r-r_H)^3 \left(r^{2}_{H}+a^2\right)^6 \Delta^{3}_\theta~ \Delta^{3},_r(r_H)~ \rho^6(r_H)} 	
 \Biggr[ 3 (-E+j \Omega+e A_0)^6  \Delta_{\theta}^6 \Sigma^6 	\nonumber\\
 & \left(r^{2}_{H}+a^2\right)^{10} \biggl\{	m^2 \left(r^{2}_{H}+a^2\right)^{2} \Delta_{\theta} - \left(J_\phi +e A_3\right)	\left(J_\phi +e A_3-1\right) \rho^2(r_H) 	\Sigma^2 \csc^2\theta \biggr\}	 \nonumber\\
 & -(-E+j \Omega+e A_0)^4 (r-r_H)  \left(r^{2}_{H}+a^2\right)^8 \Delta^{2}_\theta~ \Delta,_r(r_H) \Sigma^4			
 \Biggr( 3 m^2  \left(r^{2}_{H}+a^2\right)^2 \Delta_\theta	\nonumber\\
 &  \left(J_{\theta}^2 \Delta_{\theta}+m^2 \rho^2(r_H) 	\right)- \left(J_\phi +e A_3\right)^2 \rho^2(r_ H)  \biggl\{ J_\theta^{2}  \left(J_\phi +e A_3-1\right) \Delta_\theta -3 m^2 \rho^2(r_H) \biggr\}  \nonumber\\
 &\Sigma^2 	\csc^2\theta \Biggr) + m^2  (r-r_H)^3 \Delta^{3},_r(r_H) 
 \Biggl( 3 J_{\theta}^6 \left(r^{2}_{H}+a^2\right)^6 	 \Delta^{6}_\theta + J_{\theta}^2 	~\Delta^{2}_\theta ~ \rho^8(r_H) 	~\Sigma^4  \csc^4\theta
\nonumber\\
&	\left(J_\phi +e A_3\right)^3  \biggl\{1+2   \left(J_\phi +e A_3\right)	\biggr\}	+ \left(J_\phi +e A_3\right)^3   \rho^{10}(r_H)			~	\Sigma^4 \csc^4\theta		\biggl[ m^2 \Delta_\theta   \nonumber\\
& \biggl\{1+2   \left(J_\phi +e A_3\right)	\biggr\}		\left(r^{2}_{H}+a^2\right)^2 +3 \rho^2(r_H) 				\Sigma^2 \csc^2\theta \left(J_\phi +e A_3\right)^2  \biggr] +J_{\theta}^4 ~ \Delta_{\theta}^4~ \rho^2(r_H)  \nonumber\\ &\left(r^{2}_{H}+a^2\right)^4 \biggl\{ 3 m^2 \left(r^{2}_{H}+a^2\right)^2 \Delta_\theta + \left(J_\phi +e A_3\right) \left(2+J_\phi +e A_3\right)~ \rho^2(r_H) 	~			\Sigma^2 ~\csc^2\theta \biggr\} \Biggr)	 \nonumber\\
&	- (-E+j \Omega+e A_0)^2  \left(r^{2}_{H}+a^2\right)^2 (r-r_H)^2 \Delta^{2},_r(r_H) \Sigma^2		\Biggl( \left(1-J_\phi -e A_3\right) \left(J_\phi +e A_3\right)     \nonumber\\
& \rho^2(r_H) 		~		\Sigma^2 ~\csc^2\theta
\biggl\{ 2 J^{2}_\theta  \left(r^{2}_{H}+a^2\right)^4 \Delta_{\theta}^4+ J_{\theta}^2  \left(J_\phi +e A_3\right)^2    \left(r^{2}_{H}+a^2\right)^2 \Delta_{\theta}^2  \rho^4(r_H) 		~		\Sigma^2  \nonumber\\
& \csc^2\theta +3 \left(J_\phi +e A_3\right)^4  \rho^8(r_H) ~				\Sigma^4 ~\csc^4\theta \biggr\}   + m^2  \left(r^{2}_{H}+a^2\right)^2 \Delta_\theta 	\biggl\{ 3 J_{\theta}^4  \Delta_{\theta}^4  \left(r^{2}_{H}+a^2\right)^4 \nonumber\\
&+ \left(J_\phi +e A_3\right)^3   \left(2+J_\phi +e A_3\right)  \rho^8(r_H) ~				\Sigma^4 ~\csc^4\theta \biggr\}  \Biggr) \Biggr].
\end{align}

\section{The expressions of $\chi_1$ and $\chi_2$ given in eq. \eqref{eqn 53} }
\label{appendix B}
\begin{align}
\chi_1=& -\dfrac{3 (-E+j \Omega+e A_0)^6 m^2 \left(r^{2}_{H}+a^2\right)^6  \Sigma^6  }{(r-r_H)^3 \Delta^{3},_r(r_H)~ \rho^6(r_H)} + \dfrac{(-E+j \Omega+e A_0)^4 ~\Sigma^4}{(r-r_H)^2 ~\Delta_{\theta}^2 ~\Delta^{2},_r(r_H)~ \rho^6(r_H)} \nonumber\\ 
& \Biggl[	9 m^2  \left(r^{2}_{H}+a^2\right)^2 \Delta_{\theta} -4 \left(J_\phi +e A_3-1\right) \left(J_\phi +e A_3\right) \rho^2(r_H) 	~			\Sigma^2 ~\csc^2\theta \Biggr] \nonumber\\
& \Biggl[\Delta_\theta \left(r^{2}_{H}+a^2\right)^2 \left( J_{\theta}^2 \Delta_\theta+m^2 ~\rho^2(r_H)	\right) + \left(J_\phi +e A_3\right)^2  \rho^4(r_H) 	~			\Sigma^2 ~\csc^2\theta \Biggr] \nonumber\\ &
+m^2 \Biggl[ \dfrac{3 \left(J_\theta^{2} \Delta_\theta+m^2~ \rho^2(r_H) 	\right)^3 }{\rho^6(r_H) 	} +  \dfrac{\left(J_\phi +e A_3\right) \Sigma^2 ~\csc^2\theta
}{\left(r^{2}_{H}+a^2\right)^2 \Delta_\theta ~ \rho^2(r_H)} \Biggl\{2 J_{\theta}^2~ m^2 ~\Delta_\theta~  \rho^2(r_H)  \nonumber\\
& \bigl[ 2+ 7\left(J_\phi +e A_3\right)\bigr] + J_{\theta}^2~ \Delta_{\theta}^2 \bigl[ 4+ 5\left(J_\phi +e A_3\right)\bigr] +m^4 ~\rho^4(r_H)\bigl[ 4+ 5\left(J_\phi +e A_3\right)\bigr] \Biggr\} \nonumber\\
& +\dfrac{\rho^2(r_H) 		~		\Sigma^4 ~\csc^4\theta\left(J_\phi +e A_3\right)^3 \bigl[ 4+ 5\left(J_\phi +e A_3\right)\bigr] \left(J_\theta^{2}~ \Delta_\theta+m^2~ \rho^2(r_H) 	\right)	}{\left(r^{2}_{H}+a^2\right)^4 \Delta_{\theta}^2 } \nonumber\\
&	- \dfrac{\left(J_\phi +e A_3-4 \right)	\left(J_\phi +e A_3\right)^5 \rho^6(r_H) 				\Sigma^6 \csc^6\theta } {\left(r^{2}_{H}+a^2\right)^6 ~\Delta_{\theta}^3} \Biggr]-\dfrac{(j \Omega+e A_0-E)^2 \left(r^{2}_{H}+a^2\right)^2 \Sigma^2
 }{(r-r_H) ~\Delta,_r(r_H)~ \rho^2(r_H) }  \nonumber\\
 &  \Biggl[ 9 m^6+ m^2 \biggl\{ \dfrac{J_{\theta}^2 ~\Delta_\theta}{\rho^2(r_H) 	} + \dfrac{\left(J_\phi +e A_3\right)^2~ \rho^2(r_H)~ 	 \Sigma^2 ~\csc^2\theta}{\left(r^{2}_{H}+a^2\right)^2 ~\Delta_{\theta}} \biggr\}  \nonumber\\
 & \biggl\{ \dfrac{9 J_{\theta} \Delta_{1}}{\rho^2(r_H) } + \dfrac{\rho^2(r_H) 			~	\Sigma^2 ~\csc^2\theta \left(J_\phi +e A_3\right)\left(8+J_\phi +e A_3\right)}{\left(r^{2}_{H}+a^2\right)^2 ~\Delta_{\theta}} \biggr\}   \nonumber\\
 &   + 6 m^4 \biggl\{ \dfrac{3 J_{\theta}^2 \Delta_{\phi}}{\rho^2(r_H) }+ \dfrac{\rho^2(r_H) 			~	\Sigma^2 ~\csc^2\theta \left(J_\phi +e A_3\right)\left[1+2(J_\phi +e A_3)\right]}{\left(r^{2}_{H}+a^2\right)^2 ~\Delta_{\theta}} \biggr\} 	\nonumber\\
 & - \dfrac{4 ~\Sigma^2 ~\csc^2\theta	\left(J_\phi +e A_3-1\right)\left(J_\phi +e A_3\right)	}{\left(r^{2}_{H}+a^2\right)^6 ~\Delta_{\theta}^3 ~\rho^2(r_H)  }   \biggl\{ J_{\theta}^4 ~\Delta_{\theta}^4\left(r^{2}_{H}+a^2\right)^4  +J_{\theta}^2 ~ \Delta_{\theta}^2	~\rho^4(r_H) \nonumber\\ 
 &			\Sigma^2 ~\csc^2\theta
		\left(J_\phi +e A_3\right)^2 \left(r^{2}_{H}+a^2\right)^2 +~\rho^8(r_H) 		\Sigma^4 ~\csc^4\theta
		\left(J_\phi +e A_3\right)^4 \biggr\} \Biggr] , \\
\chi_2=& \dfrac{\rho^2(r_H) 	~			\Sigma^2 ~\csc^2\theta\left(1-J_\phi -e A_3\right) \left(J_\phi +e A_3\right)}{\left(r^{2}_{H}+a^2\right)^2 \Delta_\theta}		\Biggl[ m^2- \dfrac{ \Sigma^2 (-E+j \Omega+e A_0)^2  \left(r^{2}_{H}+a^2\right)^2}{(r-r_H) ~\Delta,_r(r_H)~ \rho^2(r_H)} \Biggr] \nonumber\\
& \Biggl[	m^2+\dfrac{J_\theta^{2} \Delta_\theta}{\rho^2(r_H)}  -  \dfrac{ \Sigma^2 (-E+j \Omega+e A_0)^2  \left(r^{2}_{H}+a^2\right)^2}{(r-r_H) ~\Delta,_r(r_H)~ \rho^2(r_H)} +\dfrac{\rho^2(r_H) 	~			\Sigma^2 ~\csc^2\theta \left(J_\phi +e A_3\right)^2}{\left(r^{2}_{H}+a^2\right)^2 \Delta_\theta}	\Biggr].
\end{align}}

\acknowledgments
The first author would like to thank the Manipur University, Canchipur for providing Non-NET fellowship. The authors also acknowledge the anonymous reviewers for valuable suggestions and comments to improve the paper.

\end{document}